\documentclass[useamsfonts]{pasj00}

\def\bmath#1{\mbox{\boldmath$#1$}}
\def\rev#1{#1}

\begin{document}
\SetRunningHead{Tomoaki Matsumoto}{Self-gravitational MHD-AMR}
\Received{2006/9/5}
\Accepted{2007/06/25}
\Published{2007/10/25}
\SetVolumeData{2007}{59}{5}

\title{Self-gravitational Magnetohydrodynamics with
  Adaptive Mesh Refinement for Protostellar Collapse}

\author{Tomoaki \textsc{Matsumoto}}

\affil{Faculty of Humanity and Environment, Hosei
 University, Fujimi, Chiyoda-ku, Tokyo 102-8160, Japan}
\email{matsu@i.hosei.ac.jp}


\maketitle

\begin{abstract}
   A new numerical code, called SFUMATO, for solving self-gravitational magnetohydrodynamics
   (MHD) problems using adaptive mesh refinement (AMR) is presented.
  A block-structured grid is adopted as the grid of the AMR hierarchy.
  The total variation diminishing (TVD) cell-centered scheme is
  adopted as the MHD solver, with hyperbolic cleaning of the divergence error of
  the magnetic field also implemented. \rev{The MHD solver exhibits a second-order accuracy in convergence tests of linearized MHD waves.}
 The self-gravity is solved using a
  multigrid method composed of (1) a full multigrid (FMG)-cycle on
  the AMR hierarchical grids, (2) a V-cycle on these grids, and (3)
 an FMG-cycle on the base grid.  The multigrid method exhibits spatial
   second-order accuracy, fast convergence, and
  scalability. The numerical fluxes are conserved by using a refluxing
  procedure in both the MHD solver and the multigrid
  method.   Several tests are performed and the results indicate that the
  solutions are consistent with previously published results.
\end{abstract}

\section{Introduction}

Protostellar collapse is one of the most important processes in star
formation; it exhibits a high dynamic range in density and spatial
dimensions.  Adaptive mesh refinement (AMR) is a powerful technique
for performing numerical simulations requiring a high dynamic range in the mesh
schemes.  Local high-resolution is realized by employing grids of
differing resolutions.  The finer grids are inserted and their
location changed according to given refinement criteria. Following the
introduction of the fundamental principles of AMR by \citet{Berger84}
and \citet{Berger89}, this technique is becoming widely used in
astrophysical simulations.

In star formation, along with self-gravity, the magnetic field also plays an 
important role.  Both self-gravity and magnetohydrodynamics (MHD)
are therefore implemented in recent AMR codes (see the review by
\cite{Klein06}).

Existing self-gravitational MHD AMR codes are generally based on
Cartesian grids; although block-structured grids tend to be most often frequently, alternative approaches also exist, e.g. RAMSES
\citep{Fromang06}. In block-structured AMR, numerical cells are
refined in the unit of the block, and a block is itself an ordinal uniform
grid.  Block-structured grids are divided into two categories:
patch-oriented grids and self-similar blocks.  In the former approach,
a block has variable number of cells, and the shape of the block is also
changed in the course of refinement.  A large block containing many cells
and a small block containing just a few cells can co-exist.  This approach
originated in the study of \citet{Berger89}, and has been adopted in many codes: e.g.,
the AMR code \rev{Orion} of Berkeley \citep{Truelove98}, Enzo \citep{Norman99},
and RIEMANN \citep{Balsara01}.  In the latter approach, all blocks
contain the same number of cells, but the physical sizes of the blocks
are different.  This approach has also been adopted by many codes: e.g.,
FLASH \citep{Fryxell00,Banerjee06}, and NIRVANA \citep{Zieglar05}.
This approach offers advantages including relatively simple algorithms
for refinement, parallelization, and vectorization.  This paper also
follows this latter approach.  In particular, this approach enables the AMR code
to implement a full multigrid (FMG)-cycle in the multigrid method
for self-gravity.  This scheme is an extension of the multigrid method
for the nested grid approach \citep{Matsumoto03} to the AMR grid.

There are several ways to treat the MHD, particularly in
the treatment of the $\nabla \cdot \bmath{B}$ term.  These approaches may be categorized as (1)
the constrained transport method with a staggered grid, (2) the
projection scheme with a Poisson solver, and (3) the eight-wave
formulation (see \cite{Toth00,Balsara04} for a comparison of these
schemes). 

Recently, \citet{Dedner02} proposed an alternative scheme for
cleaning the $\nabla \cdot \bmath{B}$ term.  This scheme has a formulation
that is similar to, but distinct from, the eight-wave formulation; two additional waves transfer
the $\nabla \cdot \bmath{B}$ error isotropically at a given speed, and
the waves decay at a given rate. As a result, the error is propagated
independently of the gas motion, and so is diluted. By contrast, using the eight-wave formulation, a single additional wave transfers the error at the gas velocity.  The error is always propagated downstream of
the gas motion, and can become stagnated at the shock wave and the center of
the collapsing cloud. The scheme of \citet{Dedner02} is
 therefore adopted here.

 The author has previously developed a self-gravitational MHD code
 with a nested grid
 \citep{Matsumoto03,Matsumoto03b,Matsumoto04,Machida04}.  In the
 nested grid approach, the grids are refined adaptively, while the
 sizes and positions of the sub-grids are fixed. The problems to which
 this method may be applied have therefore been restricted to those
 problems whose objects that require fine grid resolution are located in
 the center of the computational domain, e.g., \citet{Machida06}.
 
A self-gravitational MHD code with AMR has now, however, been
developed, and this method is presented in this paper.
In \S~\ref{sec:equations}--\ref{sec:refine},
the governing equations to be solved, the rules of discretization,
and the refinement algorithm are presented.
In \S~\ref{sec:HD}, the present methods for treating the hydrodynamics and MHD of the problem are described, while in
 \S~\ref{sec:MG}, the methods of the multigrid iteration for solving
self-gravity are described.
In \S~\ref{sec:test}, the results of several numerical tests are presented.
The paper is concluded in \S~\ref{sec:summary}.
\section{Governing Equations}
\label{sec:equations}
The equations of self-gravitational, ideal hydrodynamics and MHD 
can be expressed in 
a conservative form with a source term,
\begin{equation}
\frac{\partial \bmath{U}}{\partial t}
+ \frac{\partial \bmath{F}_x}{\partial x}
+ \frac{\partial \bmath{F}_y}{\partial y}
+ \frac{\partial \bmath{F}_z}{\partial z}
= \bmath{S} ,
\label{eq:conservative}
\end{equation}
together with Poisson's equation
\begin{equation}
\nabla^2 \Phi = 4 \pi G \rho ,
\label{eq:poisson}
\end{equation}
where $\bmath{U}$ is a vector of conservative variables, 
$\bmath{F}_x$, $\bmath{F}_y$, and $\bmath{F}_z$ are numerical fluxes,
and $\bmath{S}$ is the source term vector.  
In equation~(\ref{eq:poisson}), 
$\Phi$, $G$, and $\rho$ denote
the gravitational potential, the gravitational constant, and the density, respectively.

For ideal MHD with self-gravity, 
the vectors in equation~(\ref{eq:conservative}) are expressed by
\begin{equation}
\bmath{U} = 
\left(
\rho,
\rho v_x,
\rho v_y,
\rho v_z,
B_x,
B_y,
B_z,
\rho E
\right)^T ,
\label{eq:U}
\end{equation}
\begin{equation}
\bmath{F}_x = 
\left(
\begin{array}{c}
\rho v_x\\
\rho v_x^2 + P + \left|\bmath{B}\right|^2 / 8\pi - B_x^2/4\pi \\
\rho v_x v_y - B_x B_y/ 4\pi\\
\rho v_x v_z - B_x B_z/ 4\pi\\
0 \\
v_x B_y - v_y B_x \\
-v_z B_x + v_x B_z \\
(\rho E + P + \left|\bmath{B}\right|^2/8\pi) v_x  - B_x (\bmath{B}\cdot\bmath{v})/4\pi
\end{array}
\right) ,
\label{eq:F_x}
\end{equation}
\begin{equation}
\bmath{S} = 
\left(
0,
\rho g_x,
\rho g_y,
\rho g_z,
0,
0,
0,
\rho \bmath{g} \cdot \bmath{v}
\right)^T ,
\label{eq:S}
\end{equation}
where 
$\bmath{v} = (v_x, v_y, v_z)^{T}$ represents velocity,
$\bmath{B} = (B_x, B_y, B_z)^{T}$ represents the magnetic field,
$\bmath{g} = (g_x, g_y, g_z)^{T} = - \nabla \Phi$ represents gravity, 
$E = \left|\bmath{v}\right|^2 / 2 + (\gamma - 1)^{-1} P/\rho + \left|\bmath{B}\right|^2/8\pi\rho$ is the total energy,
and $P$ represents pressure.
The vectors $\bmath{F}_y$ and $\bmath{F}_z$ are obtained by
rotating the components in $\bmath{F}_x$ by the right-hand rule.

For ideal hydrodynamics with self-gravity, 
the governing equations are obtained by
setting $\bmath{B} = 0$ and omitting 
the 5th to the 7th components of equation~(\ref{eq:conservative}) and vectors ~(\ref{eq:U})--(\ref{eq:S}).  

Barotropic and isothermal equations of state are also implemented in
the AMR code.  In these equations of state, the component corresponding to $E$ in equation~(\ref{eq:conservative}) and vectors ~(\ref{eq:U})--(\ref{eq:S})
(the eighth component)
are excluded,  
and $P$ is expressed as a function of $\rho$.

\section{Discretization}
\label{sec:discretization}
Equations~(\ref{eq:conservative}) and (\ref{eq:poisson}) are
solved by a difference scheme based on the finite-volume approach.  The
computational domain is divided into cells, each of size $\Delta x
\times \Delta y \times \Delta z$.  Each cell is labeled by $(i,j,k)$,
the indices of the cell in the $x$, $y$, and $z$-directions, respectively. The
location of the cell center is indicated by the position vector
$\bmath{r}_{i,j,k}$.  The conservative variables $\bmath{U}$, the
source term $\bmath{S}$, and the gravitational potential $\Phi$ are
defined at the cell center, i.e., $\bmath{U}_{i,j,k} :=
\bmath{U}(\bmath{r}_{i,j,k})$, $\bmath{S}_{i,j,k} :=
\bmath{S}(\bmath{r}_{i,j,k})$, and $\Phi_{i,j,k} :=
\Phi(\bmath{r}_{i,j,k})$.  The numerical fluxes $\bmath{F}_x$,
$\bmath{F}_y$, and $\bmath{F}_z$ are defined at the cell surfaces
with normals in the $x$, $y$, and $z$-directions, respectively. For convenience,
the notation of $\bmath{F}_{x,i \pm 1/2,j,k} :=
\bmath{F}_x \left[\bmath{r}_{i,j,k} \pm (\Delta x/2)
\hat{\bmath{x}}\right]$ is introduced, where $\hat{\bmath{x}}$ denotes the unit
vector in the $x$-direction. We introduce the following notation to describe the spatial differences,
 
\begin{equation}
\partial_x Q_{i+1/2,j,k} =
\frac{Q_{i+1,j,k}-Q_{i,j,k}}{\Delta x},
\end{equation}
\begin{equation}
\partial_x Q_{i,j,k} =
\frac{Q_{i+1/2,j,k}-Q_{i-1/2,j,k}}{\Delta x},
\end{equation}
\begin{equation}
\partial_{x}^2 Q_{i,j,k} =
\frac{Q_{i+1,j,k}-2Q_{i,j,k}+Q_{i-1,j,k}}{\Delta x^2}.
\end{equation}
The differences in the $y$- and $z$-directions are expressed in a
similar manner.

\section{Grid Refinement}
\label{sec:refine}

\begin{figure}
\begin{center}
\FigureFile(80m,80mm){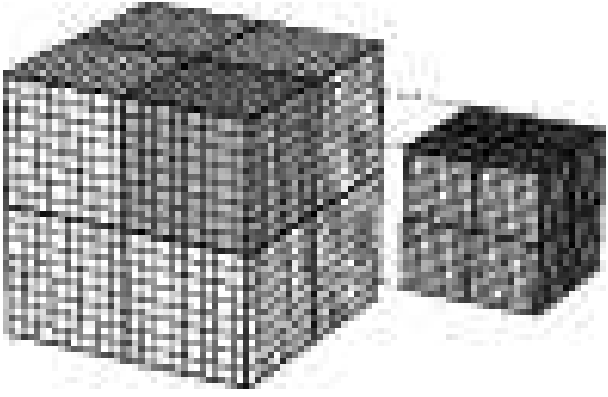}
\end{center}
\caption{ 
Schematic diagram of the grid refinement process.
\rev{	
The thick lines represent the boundaries of the blocks and the thin lines represent the boundaries of the cells.
Each block consists of $8^3$ cells in this figure.
The hatched block is refined into 8 child blocks.}
}\label{refinement.eps}
\end{figure}

A self-similar block-structured grid is adopted.  Each block consists of
$N_x \times N_y \times N_z$ cells, where $N_x$, $N_y$, and $N_z$
denote the number of cells in the $x$, $y$, and $z$-directions respectively, with all blocks having the same number of cells.  The number of cells inside a block
is fixed \rev{in course of calculation,} but the cell width differs depending on the grid-level.  The
blocks are thus self-similar. A schematic diagram of
a block-structured grid is shown in Figure~\ref{refinement.eps}. The number
of cells is set at $N_x = N_y = N_z = 8$ in this figure.

If some cells satisfy a refinement criterion, the block in which these cells lie is divided into 8 child blocks, and every cell
inside the parent block is also refined into 8 child cells (see
Fig.~\ref{refinement.eps}).  The cell width of the child cells is half
of the parent cell width, that is the refinement ratio is fixed
at two \rev{in this implementation
(in some AMRs having patch-oriented grids any refinement ratio of the power of 2 can be used).}
The coarsest grid-level is labeled $\ell = 0$ (the base
grid), and the finest grid-level is labeled $\ell =
\ell_\mathrm{\max}$.  The $\ell$-th grid-level has a $2^\ell$ times
higher spatial resolution than the coarsest grid-level.
The block-structured grid
is managed by an octree structure; the parent (coarse) block is
linked with eight fine (child) blocks.  Moreover, a block is linked
with its neighboring blocks.  These link lists are reconstructed every
time the grids are refined.

In the construction of a child block, 
the conservative variables of the parent cells $\bmath{U}^H_{I,J,K}$ are
interpolated to obtain those of the child cells $\bmath{U}^h_{i,j,k}$. For this purpose
linear interpolation with a slope limiter is adopted,
\begin{equation}
\bmath{U}^h_{i,j,k} = \bmath{U}^H_{I,J,K}
+ \overline{\nabla \bmath{U}}^H\cdot (\bmath{r}^h_{i,j,k} - \bmath{r}^H_{I,J,K}),
\label{eq:interpolation}
\end{equation}
where 
$\bmath{r}_{i,j,k}^h$ and $\bmath{r}_{I,J,K}^H$ denote the 
position vectors indicating the centers of the child and parent cells, respectively.
The gradient inside the parent cell is slope-limited according to
\begin{equation}
\overline{\nabla \bmath{U}}^H = 
\left(
\begin{array}{c}
\mathrm{minmod}\left(
\partial_x \bmath{U}^H_{I+1/2,J,K}, \partial_x \bmath{U}^H_{I-1/2,J,K}\right) \\
\mathrm{minmod}\left(
\partial_y \bmath{U}^H_{I,J+1/2,K}, \partial_y \bmath{U}^H_{I,J-1/2,K}\right) \\
\mathrm{minmod}\left(
\partial_z \bmath{U}^H_{I,J,K+1/2}, \partial_z \bmath{U}^H_{I,J,K-1/2}\right)
\end{array}
\right),
\label{eq:gradient}
\end{equation}
where 
$\mathrm{minmod}(\cdot,\cdot)$ denotes the minmod function.
Note that this interpolation conserves the conservative variables in
the refinement procedure, 
\begin{equation}
\int_{\Omega^H_{I,J,K}} \bmath{U}^H(\bmath{r}) d\bmath{r} 
= \int_{\Omega^H_{I,J,K}} \bmath{U}^h(\bmath{r}) d\bmath{r},
\end{equation}
where $\Omega^H_{I,J,K}$ denotes the zone of a parent cell whose center is located at $\bmath{r}^H_{I,J,K}$.


The refinement algorithm is based on that of \citet{Berger89}, where
grids of level $\ell$ are refined using the
following procedures to construct grids of level $\ell + 1$: 
\begin{enumerate}
\item A block of grid-level $\ell$ is marked 
  if the cells inside the block satisfy a refinement criterion.
\item If blocks of grid-level $\ell + 2$ exist, then the corresponding blocks of
  $\ell$ are also marked. 
\item  Blocks adjacent to marked blocks are also marked, so that
  the grids of level $\ell+1$ properly nest the grids of level $\ell+2$. 
\item New blocks of level $\ell + 1$ are constructed as child blocks of 
  the marked blocks.
\item Blocks of $\ell + 1$ are removed if their parent blocks are not marked.
\end{enumerate}
These procedures are called in ascending order of grid-level, from
$\ell_\mathrm{max} - 1$ to 0.

\section{Hydrodynamics and MHD}
\label{sec:HD}
\subsection{Basic Solvers}
\label{sec:basicsolver}
The governing MHD equations~(\ref{eq:conservative}), which include as a special case the hydrodynamics equations, are solved on the block-structured
grid described in \S~\ref{sec:refine}.  Methods used on ordinal
uniform grids can be applied to the block-structured grid if
boundary conditions are properly specified for each block.   A 
monotone upstream-centered scheme for conservation laws (MUSCL) approach
and predictor-corrector method are adopted here for integration with respect to
time in order to achieve a second-order accuracy in space and time
(e.g., \cite{Hirsch90}),
and an unsplit approach rather than a \rev{a fractional timestep} approach is adopted.

The numerical flux is obtained using the
linearized 
Riemann solver; the solvers are based on the
schemes of \citet{Roe81} and \citet{Fukuda99}.  For MHD,
the hyperbolic divergence cleaning of \citet{Dedner02} is adopted for
reducing $\nabla \cdot B$.
According to \citet{Dedner02}, 
equations~(\ref{eq:U})--(\ref{eq:S}) are then modified
as follows:
\begin{equation}
\bmath{U} = 
\left(
\rho,
\rho v_x,
\rho v_y,
\rho v_z,
B_x,
B_y,
B_z,
\rho E,
\psi
\right)^T ,
\label{eq:Ud}
\end{equation}
\begin{equation}
\bmath{F}_x = 
\left(
\begin{array}{c}
\rho v_x\\
\rho v_x^2 + P + \left|\bmath{B}\right|^2 / 8\pi - B_x^2/4\pi \\
\rho v_x v_y - B_x B_y/ 4\pi\\
\rho v_x v_z - B_x B_z/ 4\pi\\
\psi \\
v_x B_y - v_y B_x \\
-v_z B_x + v_x B_z \\
(\rho E + P + \left|\bmath{B}\right|^2/8\pi) v_x  - B_x (\bmath{B}\cdot\bmath{v})/4\pi\\
c_h^2 B_x
\end{array}
\right) ,
\label{eq:F_xd}
\end{equation}
\begin{equation}
\bmath{S} = 
\left(
\begin{array}{c}
0 \\
\rho g_x - (\nabla \cdot \bmath{B}) B_x /4\pi\\
\rho g_y - (\nabla \cdot \bmath{B}) B_y /4\pi\\
\rho g_z - (\nabla \cdot \bmath{B}) B_z /4\pi\\
0 \\
0 \\
0 \\
\rho (\bmath{g} \cdot \bmath{v}) - \bmath{B} \cdot (\nabla \psi)/4\pi \\
-(c_h/c_p)^2 \psi
\end{array}
\right) ,
\label{eq:Sd}
\end{equation}
where $\psi$ is a scalar potential propagating a divergence error,
$c_h$ is the wave speed, and $c_p$ is the damping rate of the wave.  Note
that this modification requires additional components in the basic
equation, and a total of nine waves are solved for. This approach is similar to
the eight-wave formulation.  In the method of \citet{Dedner02}, two
additional waves transfer the $\nabla \cdot \bmath{B}$ error
isotropically at a speed of $c_h$, and the waves decay at a rate
of $c_p$. Similarly, in the eight-wave formulation, an additional wave
transfers the divergence error at a gas velocity of $\bmath{v}$.

In the predictor step, $\bmath{U}^n_{i,j,k}$ is updated to
$\bmath{U}^{n+1/2}_{i,j,k}$ by a half time step, i.e., 
\begin{eqnarray}
\bmath{U}^{n+1/2}_{i,j,k} 
&=& \bmath{U}^{n}_{i,j,k} \nonumber\\
&-& \frac{\Delta t}{2}
\left(
\partial_x \bmath{F}_{x, i,j,k}^{n} 
+\partial_y \bmath{F}_{y, i,j,k}^{n}
+\partial_z \bmath{F}_{z, i,j,k}^{n}
\right.\nonumber\\
&-& \bmath{S}_{i,j,k}^{n,*} 
\left.
\right),
\label{eq:predictor}
\end{eqnarray}
where the superscript $n$ denotes the time level,
and $\Delta t = t^{n+1} - t^n$.
The numerical flux $\bmath{F}^n$, which is defined at the cell
boundaries, \rev{is calculated using the primitive variables,}
\begin{equation}
  \bmath{Q} = \left(
\rho, v_x, v_y, v_z, B_x, B_y, B_z, P, \psi
\right)^T. \label{eq:primitive_variables}
\end{equation}
\rev{ 
The numerical flux has second-order spatial accuracy due to the MUSCL
extrapolation of the amplitudes of the eigenmode of the MHD waves,
$\bmath{L} \bmath{U}$, where $\bmath{L}$ denotes the 
matrix of the left eigenvector
(see the appendix in \cite{Fukuda99} for details)
\footnote{\rev{
For the case of the hydrodynamics, the MUSCL extrapolation is applied to the
 primitive variables, $\bmath{Q}$ in the predictor and corrector steps
in this implementation.}}. }

The source term $\bmath{S}_{i,j,k}^{n,*}$ is calculated using
\begin{eqnarray}
\lefteqn{
\bmath{S}_{i,j,k}^{n,*} } \\
&=&
\left(
\begin{array}{c}
0 \\
\rho_{i,j,k}^n g_{x,i,j,k}^{n-1/2} - (\nabla \cdot \bmath{B})_{i,j,k}^n B_{x,i,j,k}^n /4\pi\\
\rho_{i,j,k}^n g_{y,i,j,k}^{n-1/2} - (\nabla \cdot \bmath{B})_{i,j,k}^n B_{y,i,j,k}^n /4\pi\\
\rho_{i,j,k}^n g_{z,i,j,k}^{n-1/2} - (\nabla \cdot \bmath{B})_{i,j,k}^n B_{z,i,j,k}^n /4\pi\\
0 \\
0 \\
0 \\
\rho_{i,j,k}^n (\bmath{g}_{i,j,k}^{n-1/2} \cdot \bmath{v}_{i,j,k}^n)
- \bmath{B}_{i,j,k}^n \cdot (\nabla \psi)_{i,j,k}^n /4\pi\\
0
\end{array}
\right) ,\nonumber
\label{eq:predictor-S}
\end{eqnarray}
where $(\nabla \cdot \bmath{B})_{i,j,k} = \partial_x B_{x,i,j,k}
+ \partial_y B_{y,i,j,k} + \partial_z B_{z,i,j,k}$, and
the magnetic field defined at the cell surfaces,
$B_{x,i\pm1/2,j,k}$, is obtained from the ninth component of $\bmath{F}_{x,i\pm1/2,j,k}$.
Similarly, $(\nabla \psi)_{i,j,k} \rev{= 
(\partial_x \psi_{i,j,k}, \partial_y \psi_{i,j,k}, \partial_z \psi_{i,j,k})^T}$ 
is obtained from the fifth to seventh
components of $\bmath{F}_{x,i\pm1/2,j,k}$, $\bmath{F}_{y,i,j\pm1/2,k}$, and $\bmath{F}_{z,i,j,k\pm1/2}$.
The ninth component of the source term~(\ref{eq:Sd}) is evaluated separately by operator
splitting, in which the formal solution is used as follows,
\begin{equation}
\psi^{n+1/2} = \psi^{n+1/2,*} \exp \left[
-\frac{\Delta t}{2}
\left(\frac{c_h}{c_p}\right)^2 \right],
\end{equation}
where
$\psi^{n+1/2,n*}$ is the ninth component of $\bmath{U}^{n+1/2}$ solved
by equation~(\ref{eq:predictor}).
The free parameters $c_h$ and $c_p$ are related to the time-marching,
and are described in \S~\ref{sec:time-marching}.

Note that gravity $\bmath{g}_{i,j,k}^{n-1/2}$ lags by half a time step. This slight lagging is expected to have a negligible
effect on the accuracy in the predictor step, and the gravity in 
the previous corrector step can be reused in this predictor step
\citep{Truelove98}.  This avoids an additional call of the multigrid method
to solve Poisson's equation, significantly reducing the computational
costs of this method.

In the corrector step, 
a spatially second-order numerical flux
$\bmath{F}^{n+1/2}$ is obtained by applying MUSCL extrapolation to
\rev{the amplitudes of the eigenmodes,} which are converted from
$\bmath{U}^{n+1/2}$.
Using this flux,
$\bmath{U}^{n}_{i,j,k}$ is updated to $\bmath{U}^{n+1}_{i,j,k}$ by a full time
step, 
\begin{eqnarray}
\bmath{U}^{n+1}_{i,j,k} &=& \bmath{U}^{n}_{i,j,k} \nonumber\\
&-& \Delta t
\left(
 \partial_x \bmath{F}_{x, i,j,k}^{n+1/2} 
+\partial_y \bmath{F}_{y, i,j,k}^{n+1/2}
+\partial_z \bmath{F}_{z, i,j,k}^{n+1/2} \right. \nonumber\\
&-&\left. \bmath{S}_{i,j,k}^{n+1/2}\right).
\label{eq:corrector}
\end{eqnarray}
The source term in the corrector step is estimated at the time level $t = t^{n+1/2}$,
\begin{eqnarray}
\lefteqn{
\bmath{S}_{i,j,k}^{n+1/2}
=}  \\
&&\left(
\begin{array}{c}
0 \\
\rho_{i,j,k}^{n+1/2} g_{x,i,j,k}^{n+1/2} - (\nabla \cdot \bmath{B})_{i,j,k}^{n+1/2} B_{x,i,j,k}^{n+1/2} /4\pi\\
\rho_{i,j,k}^{n+1/2} g_{y,i,j,k}^{n+1/2} - (\nabla \cdot \bmath{B})_{i,j,k}^{n+1/2} B_{y,i,j,k}^{n+1/2} /4\pi\\
\rho_{i,j,k}^{n+1/2} g_{z,i,j,k}^{n+1/2} - (\nabla \cdot \bmath{B})_{i,j,k}^{n+1/2} B_{z,i,j,k}^{n+1/2} /4\pi\\
0 \\
0 \\
0 \\
\rho_{i,j,k}^{n+1/2} (\bmath{g}_{i,j,k}^{n+1/2} \cdot \bmath{v}_{i,j,k}^{n+1/2})
- \bmath{B}_{i,j,k}^{n+1/2} \cdot (\nabla \psi)_{i,j,k}^{n+1/2} /4\pi\\
0
\end{array}
\right) ,\nonumber
\label{eq:corrector-S}
\end{eqnarray}
where $\rho^{n+1/2}_{i,j,k}$ and $\bmath{v}_{i,j,k}^{n+1/2}$ are
obtained from $\bmath{U}^{n+1/2}_{i,j,k}$, and $\bmath{g}_{i,j,k}^{n+1/2}$ is
obtained by solving Poisson's equation~(\ref{eq:poisson}) at the half 
time step as follows,
\begin{equation}
\left(\nabla^2 \Phi \right)^{n+1/2}_{i,j,k} = 4 \pi G \rho^{n+1/2}_{i,j,k}.
\label{eq:poisson_ijk}
\end{equation}
Poisson's equation is solved by the multigrid method, as shown in
\S~\ref{sec:MG}, and the multigrid method prepares not only the
cell-centered $\Phi$, but also the cell-surfaced $\bmath{g}$.  
Gravity at the cell center $\bmath{g}_{i,j,k}^{n+1/2}$ is obtained by
averaging the values of gravity at the cell surfaces,
\begin{equation}
\bmath{g}_{i,j,k}^{n+1/2}  = 
\frac{1}{2}
\left(
\begin{array}{c}
g^{n+1/2}_{x, i+1/2,j,k} + g^{n+1/2}_{x, i-1/2,j,k} \\
g^{n+1/2}_{y, i,j+1/2,k} + g^{n+1/2}_{y, i,j-1/2,k} \\
g^{n+1/2}_{z, i,j,k+1/2} + g^{n+1/2}_{z, i,j,k-1/2}
\end{array}
\right) .
\label{eq:gravity}
\end{equation}
Note that $\bmath{g}$ at a cell surface is obtained by the 
difference of the cell-centered $\Phi$ (see
equations~[\ref{eq:gx_discrete}]--[\ref{eq:gz_discrete}]). 
Equation~(\ref{eq:gravity}) therefore coincides with the 
central difference of $\Phi$ as far as the inside of a block is concerned.

Similar to the predictor step, the ninth component of the source term is
evaluated by operator splitting,
\begin{equation}
\psi^{n+1} = \psi^{n+1,*} \exp \left[-\Delta t
  \left(\frac{c_h}{c_p}\right)^2 \right], 
\end{equation}
where
$\psi^{n+1,*}$ is the ninth component of $\bmath{U}^{n+1}$ obtained by
equation~(\ref{eq:corrector}).

\subsection{Time Marching}
\label{sec:time-marching}
The time-marching represented by equations~(\ref{eq:predictor}) and
(\ref{eq:corrector}) proceeds in units at the grid-level.  The code is
equipped with two modes of time-marching: an adaptive and a
synchronous time-step mode.  Which mode is used depends on whether the
gas is non-self-gravitational or self-gravitational.  The adaptive
time-step mode is appropriate for non-self-gravitational gases, and is
based on the method of \citet{Berger89}. In this mode, a coarser grid
has a longer time step than a finer grid.  In contrast, the
synchronous mode is appropriate for self-gravitational gases, and
every grid-level has the same time step.  This is because evolution on
the fine grid affects the detached coarse grid immediately due to 
self-gravity, and so the same time step must to be chosen for every
grid-level (see discussion of \cite{Truelove98}).  Note that the
adaptive time-step mode could also be used for a self-gravitational
gas at the expense of the first-order temporal
accuracy of the scheme. 

\begin{figure}
\begin{center}
\FigureFile(80m,80mm){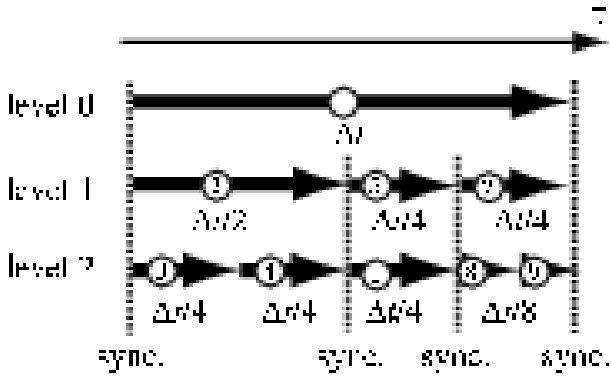}
\end{center}
\caption{ 
Adaptive time-step scheme.
The thick arrows denote time steps, while the associated numbers,
\textcircled{\tiny 1}, \textcircled{\tiny 2}, ..., \textcircled{\tiny 9},
indicate the order of time-marching.
}\label{time_marching.eps}
\end{figure}

Figure~\ref{time_marching.eps} shows the order in which the
grid-levels proceed for the adaptive time-step mode schematically.
The numbers associated with the thick arrows denote the order of 
time-marching.  Coarser grid-levels precede finer
grid-levels.  The fine grid-level undergoes several sub-cycles
until the time level of the fine grid-level is synchronized with that
of the coarse grid-level.  The time step of a finer grid-level, $\Delta
t^h$, is given by
\begin{equation}
\Delta t^h = \Delta t^H 2^{-n} ,
\end{equation}
\begin{equation}
n = \min\left\{ m \in \mathbb{N}
\;| \; \Delta t^H 2^{-m} \leq \Delta t^h_\mathrm{CFL} \right\}.
\end{equation}
where $\Delta t^h_\mathrm{CFL}$ denotes the time step calculated
directly by the CFL condition at the fine grid-level, 
and $\Delta t^H$ denotes the time step
at the coarser grid-level.
Note that $\Delta t^h$ is fixed at $\Delta t^H/r$ in
\citet{Berger89}, where $r$ denotes a refinement ratio. 
On the other hand, in the method presented here
$\Delta t^h$ can be equal to $\Delta t^H/2^n$ for $n = 0, 1, 2, \cdots$,
if $\Delta t^h$  satisfies the CFL condition.
In the synchronous time-step mode, a common time step $\Delta t$ is
used at all grid-levels, and is given by
\begin{equation} 
\Delta t =
\min_{0 \le \ell \le \ell_{\max}}
\left\{\Delta t^\ell_\mathrm{CFL}\right\},
\end{equation}
where $\Delta t^\ell_\mathrm{CFL}$ denotes the time step calculated by
the CFL condition at grid-level $\ell$.

The solver described in \S~\ref{sec:basicsolver} includes two parameters,
$c_h$ and $c_p$, and these are related to the mode of the time-marching.
The wave speed $c_h$ is obtained as
\begin{equation}
c_h = \mathrm{CFL}\frac{h}{\Delta t^{\ell=0}},
\end{equation}
where CFL denotes the CFL number, and $\Delta t^{\ell=0}$ denotes the
time step at grid-level $\ell=0$.
For the adaptive time-step mode, 
\begin{equation}
h = \min_{\ell=0}\left\{\Delta x^\ell,\Delta y^\ell, \Delta z^\ell\right\}.
\end{equation}
while for the synchronous time-step mode,
\begin{equation}
 h = \min_{\ell=\ell_\mathrm{max}}\left\{\Delta x^\ell,\Delta y^\ell, \Delta z^\ell\right\},
\end{equation}
In both cases, $c_h$ is constant across all the grid-levels, 
and satisfies the CFL condition at every grid-level.

The damping rate $c_p$ is obtained from $c_h$,
\begin{equation}
  c_p^2 = 0.18 L c_h,
\end{equation}
where $L$ is a scale length of a problem, and 
the coefficient of 0.18 is chosen according to \citet{Dedner02}.
\rev{
This coefficient is a free parameter specifying the ratio of the damping
rate and propagation speed of $\nabla \cdot \bmath{B}$.
We confirmed that this selection worked well in many calculations (e.g., 
\cite{Machida05a, Machida05b}) and in the numerical tests presented in this paper.
}

\subsection{Boundary Condition for Ghost Cells}

\begin{figure*}
\begin{center}
\FigureFile(160m,160mm){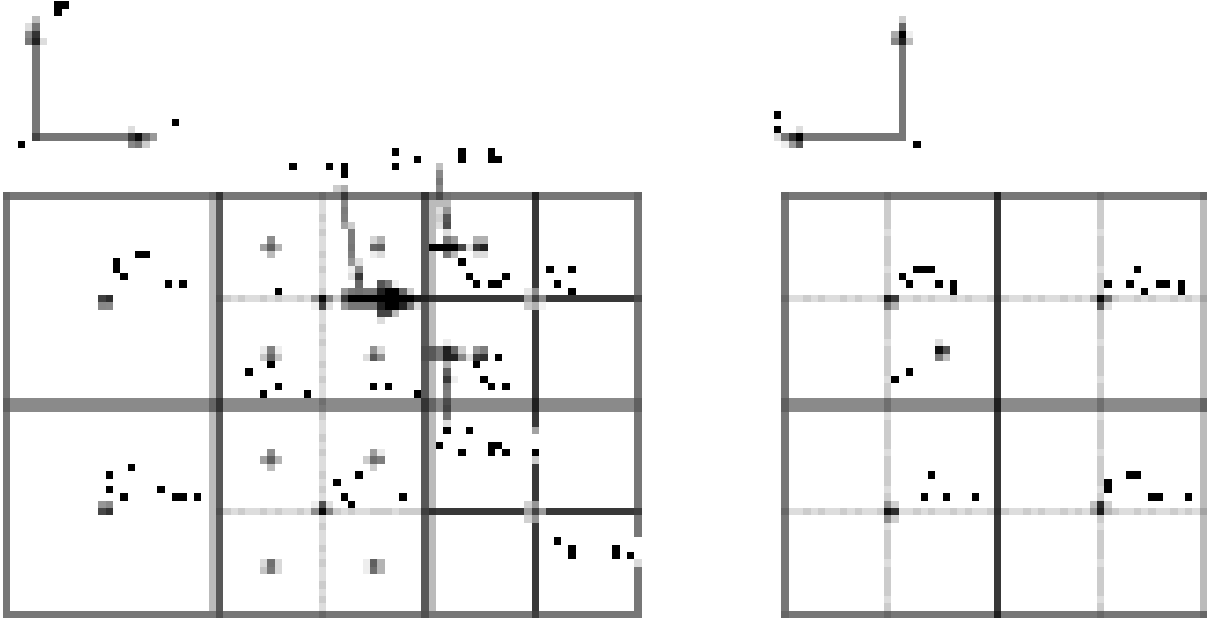}
\end{center}
\caption{ 
The interface between
a coarse block and a fine block.
The grid points of the ghost cells for the fine block are
denoted by gray circles, while the real grid points are denoted by
filled circles.  The arrows denote the numerical fluxes through the
interface between the fine and coarse blocks.
}\label{hd_grid_boundary.eps}
\end{figure*}

Each block has ghost cells overlapping with the adjacent blocks.
Figure~\ref{hd_grid_boundary.eps} shows the interface between 
the coarse and fine blocks in the $x$-direction.
Two ghost cells are prepared in each direction,
as indicated by the gray circles in the figure,
since the MUSCL extrapolation requires two cells 
on both the left and right sides of a cell boundary 
where numerical flux is obtained.

A boundary condition is imposed on the ghost cells each time before
the predictor and corrector steps proceed.  When a block is adjacent
to another block of the same grid-level, data is exchanged
between the two blocks by a simple copy.  When a block is adjacent to
a coarse block, data is interpolated spatially.  In addition,
data is interpolated temporally if an adaptive time-step
mode is used.  The procedure for the boundary conditions is
as follows:
\begin{enumerate}
\item Adjacent coarse blocks are identified.
\item Conservative variables on the coarse cells overlapping
  the ghost cells of a fine block are interpolated
  temporally to coincide with the time level of the fine grid-level.
  This procedure is omitted if the synchronous time-step mode is used.
\item The temporally interpolated variables are converted into 
  primitive variables.  They are interpolated spatially to coincide
  with the grid points of the ghost cells of the fine grid-level. The
  interpolated variables are copied into the ghost cells.
\item Adjacent blocks in the same grid-level are identified.
\item The variables of  cells overlapping the ghost cells are
  simply copied into the ghost cells.
\end{enumerate}

In the temporal interpolation, 
the quadratic interpolation is performed on $\bmath{U}^{H,n}$,
$\bmath{U}^{H,n+1/2}$, and $\bmath{U}^{H,n+1}$.
Note that $\bmath{U}^{H,n}$ and $\bmath{U}^{H,n+1}$ have of second-order temporal accuracy while $\bmath{U}^{H,n+1/2}$ is of first-order temporal accuracy.
Using these variables, $\bmath{U}^H(t)$ in $t^n \le t \le t^{n+1}$
is interpolated as,
\begin{eqnarray}
\bmath{U}^H(t) &=& \left( \bmath{U}^{H,n+1} - \bmath{U}^{(1),H,n+1} \right) 
\left( \frac{t-t^{n}}{t^{n+1} - t^{n}} \right)^2 \nonumber\\
&+& \frac{ (t^{n+1} - t) \bmath{U}^{H,n} + (t-t^{n})
  \bmath{U}^{(1),H,n+1}}{t^{n+1} - t^{n}} ,
\end{eqnarray}
where $\bmath{U}^{(1),H,n+1}$ denotes the conservative variables of
first-order accuracy at the time level $t^{n+1}$, defined as,
\begin{equation}
\bmath{U}^{(1),H,n+1} = 2 \bmath{U}^{H,n+1/2} -  \bmath{U}^{H,n}.
\end{equation}

A slope-limited gradient is adopted for the spatial interpolation.
In Figure~\ref{hd_grid_boundary.eps}, 
the gradient of the primitive variables 
inside the cell $(I,J,K)$ is evaluated using 
\begin{equation}
\nabla \bmath{Q}^H = 
\left(
\begin{array}{c}
\partial_x \bmath{Q}^H_{I-1/2,J,K} \\
\mathrm{minmod}\left(
\partial_y \bmath{Q}^H_{I,J+1/2,K}, \partial_y \bmath{Q}^H_{I,J-1/2,K}\right) \\
\mathrm{minmod}\left(
\partial_z \bmath{Q}^H_{I,J,K+1/2}, \partial_z \bmath{Q}^H_{I,J,K-1/2}\right)
\end{array}
\right).
\label{eq:gradient2}
\end{equation}
It may be noted that equation~(\ref{eq:gradient2}) describes a simple interpolation in
the normal direction, with a slope-limited interpolation in the
transverse directions.

\subsection{Refluxing at the Interfaces}
\label{sec:reflux}
Refluxing was introduced by \citet{Berger89}, and has often been used in standard implementations of AMRs. This technique maintains consistency between the fine and coarse grid-levels during time-marching, and ensures that conservation laws, e.g., mass conservation, are satisfied.

\begin{figure}
\begin{center}
\FigureFile(80m,80mm){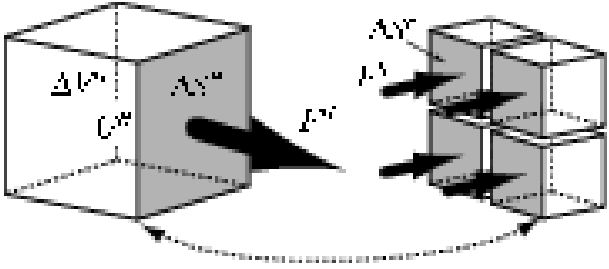}
\end{center}
\caption{ 
Refluxing procedure at the interface between
fine and coarse cells.
}\label{boundary_flux.eps}
\end{figure}

As shown in Figure~\ref{hd_grid_boundary.eps}, $\bmath{U}^H_{I,J,K}$
is updated using $\bmath{F}^H_{x,I+1/2,J,K}$, while
$\bmath{U}^h_{i,\hat{j},\hat{k}}$ is updated using
$\bmath{F}^{h,n}_{x,i-1/2,\hat{j},\hat{k}}$ 
through several sub-cycles until 
the fine grid-level catches up with the coarse grid-level,
where $\hat{j} \in [j, j+1]$,
$\hat{k} \in [k, k+1]$, and $n$ denotes the index of the sub-cycles of the
fine grid-level.
Obviously,
$\bmath{F}^H_{x,I+1/2,J,K} \Delta S^H \Delta t^H$ should equal 
$\Sigma_{\hat{j},\hat{k},n}\bmath{F}^{h,n}_{x,i-1/2,\hat{j},\hat{k}}
\Delta S^h \Delta t^{h,n} $  
to ensure that the conservation laws are satisfied, where $\Delta S^H = \Delta y^H \Delta
z^H$ and $\Delta S^h = \Delta y^h \Delta z^h$.
In refluxing, $\bmath{U}^H_{I,J,K}$ is re-calculated, taking
account of the difference between $\bmath{F}^H_{x,I+1/2,J,K}\Delta S^H \Delta t^H$  
and
$\Sigma_{\hat{j},\hat{k},n}\bmath{F}^{h,n}_{x,i-1/2,\hat{j},\hat{k}}
\Delta S^h \Delta t^{h,n}$. 
More explicitly, as indicated in Figure~\ref{boundary_flux.eps}, 
the refluxing procedure 
updates $\bmath{U}^{H,*}$ to $\bmath{U}^H$  according to
\begin{eqnarray}
\bmath{U}^H &=& \bmath{U}^{H,*} \nonumber\\
&-&
\frac{1}{\Delta V^H}
\Bigl(
\sum_n \sum _\mathrm{surface} \bmath{F}_x^{h,n}  \Delta t^{h,n} \Delta S^h 
 \nonumber\\
&-&  \bmath{F}_x^H  \Delta t^H \Delta S^H
\Bigr)
,
\end{eqnarray}
where
\begin{equation}
\sum_n \Delta t^{h, n} = \Delta t^H ,
\end{equation}
\begin{equation}
\sum_\mathrm{surface} \Delta S^h = \Delta S^H ,
\end{equation}
and $\Delta V^H$ denotes the volume of the coarse cell.

\section{Multigrid Method for Self-gravity}
\label{sec:MG}

\subsection{Multigrid Cycles}

The multigrid method is widely used in many AMRs for solving Poisson's
equation.  In many AMRs, a solution converges by means of the
V-cycle on the grids of AMR hierarchy, and the FMG-cycle on a uniform
base grid.  By contrast, the multigrid method presented here
uses not only V-cycles, but also FMG-cycles on the grids of the
AMR hierarchy.  An FMG-cycle on the hierarchical grids is also
implemented in the approach of \citet{Matsumoto03}, and the same strategy is
adopted here.

\begin{figure*}[t]
\begin{center}
\FigureFile(160m,160mm){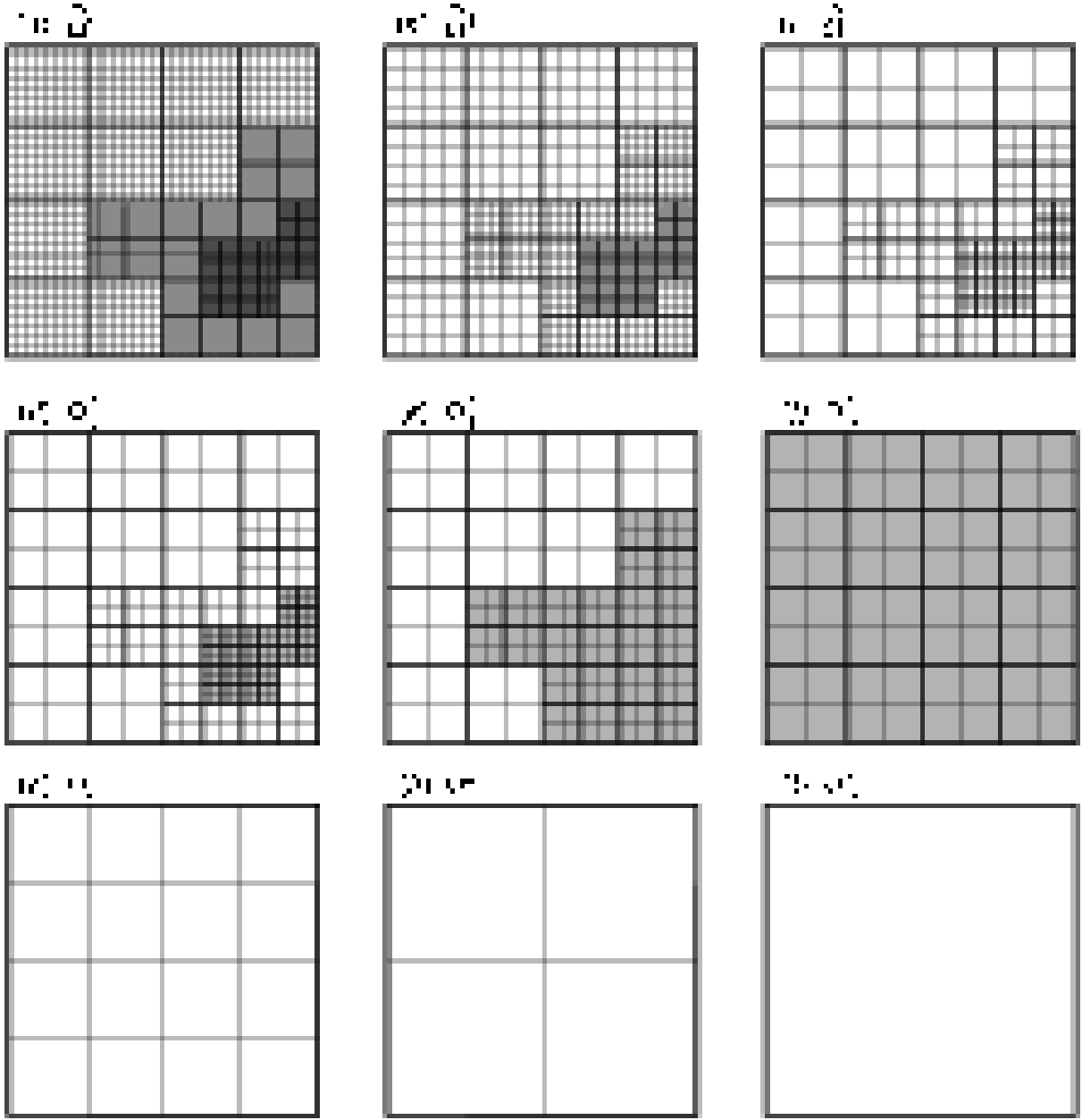}
\end{center}
\caption{ 
Schematic diagram of the coarsening of grids in the multigrid method.
The cell-boundaries and block-boundaries are denoted by thin and thick lines, respectively.
({\it a}--{\it c}) 
Coarsening of grids in the FMG-cycle on the AMR hierarchy.
The number of the cells per block decreases up to $2^3$.
({\it d}--{\it f}) 
V-cycle on the AMR hierarchy. 
The solution converges sequentially on the hatched blocks.  
({\it g}--{\it i})
Coarsening of grids in the FMG-cycle on the uniform base grid.
}\label{fmg_grid.eps}
\end{figure*}

Figure~\ref{fmg_grid.eps} shows the grids used in the multigrid cycles
for the case $N_x = N_y = N_z = 8$.  
For convenience, a grid is labeled by
\begin{equation}
\Omega_\ell^m \quad \mbox{with} \quad 
\ell \in \left[ 0, \ell_\mathrm{max} \right]
\quad \mbox{and} \quad  
m \in \left[ 0, m_\mathrm{max} \right],
\end{equation}
where $\ell$ and $m$ denote the grid-level of
the AMR hierarchy and
the coarsening level, respectively. 
For example, 
the hatched grid in
Figure~\ref{fmg_grid.eps}{\it e} is labeled as $\Omega_1^2$, where
the grid of $\ell=1$ is coarsened by a factor $2^2$.
The coarsening level of all the grids
shown in Figure~\ref{fmg_grid.eps}{\it a} is $m=0$, while that of the grids shown in
Figure~\ref{fmg_grid.eps}{\it b} is $m=1$, irrespective of 
the grid-level $\ell$.
In addition a composite grid is defined as
\begin{equation}
\hat{\Omega}^m
:= \Omega_0^m \cup \cdots \cup \Omega_{\ell_\mathrm{max}}^m .
\end{equation}
For example, the composite grids shown
in Figures~\ref{fmg_grid.eps}{\it a}, \ref{fmg_grid.eps}{\it b}, and
\ref{fmg_grid.eps}{\it c} are expressed as $\hat{\Omega}^0$,
$\hat{\Omega}^1$, and $\hat{\Omega}^2$, respectively.

\begin{figure*}[t]
\begin{center}
\FigureFile(140m,140mm){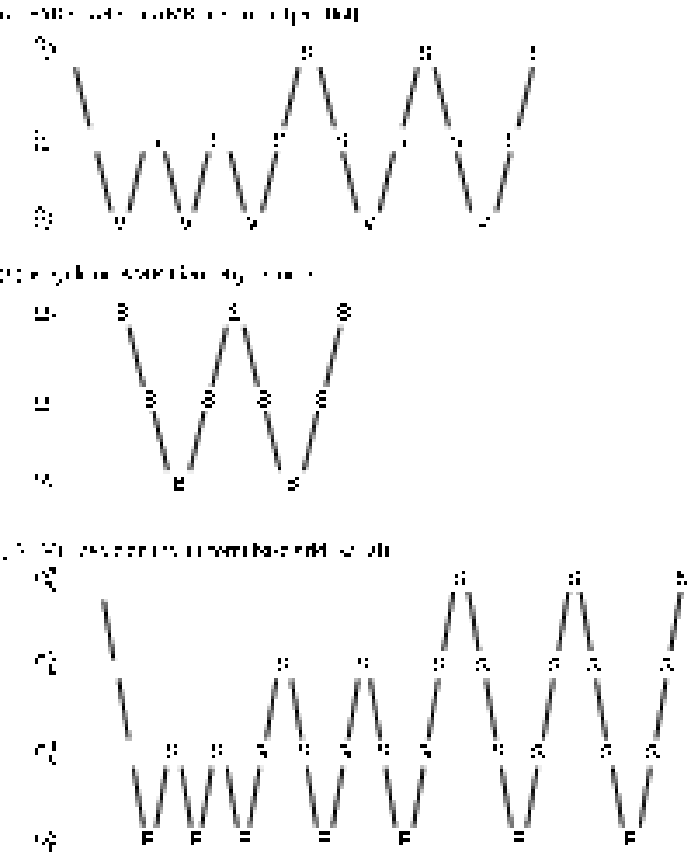}
\end{center}
\caption{ 
Schematic diagram of the multigrid cycles.
The lines pointing diagonally downwards from left to right denote the restriction operators, while the lines pointing diagonally upwards from left to right denote the prolongation operators.
The {\sf S} symbols denote the smoothing operators.
The {\sf V} symbols in panel {\it a} denote
the V-cycles on the AMR hierarchy shown in panel {\it b}.
The {\sf B} symbols in panel {\it b} denote
the FMG-cycles on the uniform base grid shown in panel {\it c}.
At the point denoted by {\sf E}, an exact solution is obtained
according to the boundary conditions.
}\label{fmg.eps}
\end{figure*}

An overview of the cycles used by the present numerical method is now presented.
First, the solution 
on $\hat{\Omega}^m$ for 
$0 \le m \le \hat{m}_\mathrm{max}$ converges under FMG-cycles,
as shown in 
Figures~\ref{fmg_grid.eps}{\it a}-\ref{fmg_grid.eps}{\it c} 
and \ref{fmg.eps}{\it a}.
The maximum coarsening level in the FMG-cycle is given by
$\hat{m}_\mathrm{max} = \log_2 \min\left(N_x,N_y,N_z\right)-1$.
In other words, the grid is coarsened until the number of the cells per block
is decreased up to two in at least one direction.
At the bottom of the FMG-cycle, which is marked by {\sf V} in 
Figure~\ref{fmg.eps}{\it a}, 
the solution then converges on $\Omega_\ell^m$ for
$0 \le \ell \le \ell_\mathrm{max}$ and $m=\hat{m}_\mathrm{max}$
under V-cycles  
(the hatched grids in Figures~\ref{fmg_grid.eps}{\it
  d}-\ref{fmg_grid.eps}{\it f}).
Typically, 
only one iteration of the V-cycle is sufficient, although, for reference,
two cycles of the V-cycle are illustrated in Figure~\ref{fmg.eps}{\it b}.  
At the bottom of the V-cycle, which is marked by {\sf B} in 
Figure~\ref{fmg.eps}{\it b}, the solution 
on the coarsened base grid $\Omega_0^m$ for $m \ge
\hat{m}_\mathrm{max}$ converges under FMG-cycles,
as shown in 
Figures~\ref{fmg_grid.eps}{\it g}-\ref{fmg_grid.eps}{\it i} 
and \ref{fmg.eps}{\it c}.
Finally, at the bottom of the FMG-cycle on the base grid, which is
marked by {\sf E} in Figure~\ref{fmg.eps}{\it c}, 
an exact solution is given according to the boundary conditions,
since there is only one cell in the computational domain.

Note that the computation at $\hat{\Omega}^0$ of the FMG-cycle accounts for
most of the computational time of the multigrid method, because the number of cells
in $\hat{\Omega}^0$ dominates all others.  The number of cells
decreases by a factor of 1/8 every time the grids are coarsened.  This
indicates that the computation on $\hat{\Omega}^0$ dominates the
overall computational cost, and this scheme is scalable to the number
of cells in a similar way as the hydrodynamics scheme (see also \cite{Matsumoto03}).

\subsection{Smoothing}
\label{sec:smoothing}
The red-black Gauss-Seidel iteration is
adopted as a smoothing procedure.  Conventional red-black Gauss-Seidel
iteration can be applied only to a uniform grid. This iteration scheme is therefore
modified so that it can be applied to composite grids, in which grids
with different resolutions co-exist.

The discretization of Poisson's equation on the composite grids is now
described.  Poisson's equation can be expressed as a set of two equations,
\begin{equation}
 \nabla \cdot \bmath{g} = -4 \pi G \rho,
\end{equation}
\begin{equation}
 \bmath{g} =  -\nabla \Phi.
\end{equation}
These equations are discretized as
\begin{equation}
\partial_x g_{x, i,j,k}
+\partial_y g_{y, i,j,k}
+\partial_z g_{z, i,j,k}
= -4 \pi G \rho_{i,j,k},
\label{eq:poisson_discrete}
\end{equation}
\begin{equation}
g_{x, i+1/2,j,k} = -\partial_x \Phi_{i+1/2,j,k},
\label{eq:gx_discrete}
\end{equation}
\begin{equation}
g_{y, i,j+1/2,k} = -\partial_y \Phi_{i,j+1/2,k},
\label{eq:gy_discrete}
\end{equation}
\begin{equation}
g_{z, i,j,k+1/2} = -\partial_z \Phi_{i,j,k+1/2},
\label{eq:gz_discrete}
\end{equation}
where $g_{x,i+1/2,j,k}$, $g_{y,i,j+1/2,k}$, and $g_{z,i,j,k+1/2}$ are
the components of gravity defined on the cell surfaces.

\begin{figure*}
\begin{center}
\FigureFile(140m,140mm){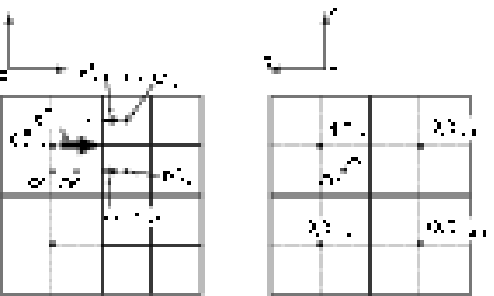}
\end{center}
\caption{ 
Interface between the fine and coarse cells for the multigrid method.
}
\label{fmg_grid_boundary.eps}
\end{figure*}

Considering the fine cells adjacent to the 
interface between the fine and coarse cells 
in Figure~\ref{fmg_grid_boundary.eps},
the potential on the ghost cell, $\Phi^{B}$ is required in
order to obtain $g^h_{x,i-1/2,j,k}$.
The potential on the ghost cell is given by,
\begin{equation}
\Phi^{B} =  \frac{10  \Phi^h_{i,j,k} + 8 \Phi^{*} - 3 \Phi^h_{i+1,j,k}}{15},
\label{eq:fmg_boundary1}
\end{equation}
\begin{eqnarray}
\lefteqn{\Phi^{*} = } \nonumber\\
&& \frac{9 \Phi^H_{I,J,K} + 3\left(\Phi^H_{I,J,K-1} +
  \Phi^H_{I,J-1,K}\right) + \Phi^H_{I,J-1,K-1}}{16},
\label{eq:fmg_boundary2}
\end{eqnarray}
where $\Phi^{B}$ is obtained by quadratic interpolation in the
$x$-direction, and $\Phi^{*}$ is obtained by bilinear
interpolation in the $y-z$ plane. The quadratic interpolation in
the direction normal to the interface satisfies the necessary conditions
for so-called conservative interpolation (see
\cite{Trottenberg01}).  Using this procedure, the components of gravity of the fine
cells adjacent to coarse cells are obtained.

For the coarse cells adjacent to fine cells, the gravity at the
interface is given by summing the gravity on the corresponding cell
surfaces of the fine cells.  For the example of Figure~\ref{fmg_grid_boundary.eps},
such a coarse gravity is given by,
\begin{equation}
g^{H}_{x,I+1/2,J,K} = \frac{1}{4}\sum_{\hat{j}=j}^{j+1} \sum_{\hat{k}=k}^{k+1}  g^h_{x,i-1/2,\hat{j},\hat{k}}
\label{eq:reflux_g}
\end{equation}
instead of $g^H_{x,I+1/2,J,K} = \partial_x \Phi^H _{I+1/2,J,K}$.  This
is a similar strategy to the refluxing described in \S~\ref{sec:reflux},
and ensures that the solution satisfies Gauss's theorem; when the
normal component of gravity is summed over the surfaces of any
cell, the sum equals the mass contained by the cell multiplied by
$4 \pi G$,
\begin{equation}
\sum _\mathrm{surface} \bmath{g} \cdot \Delta\bmath{S}
 = - 4 \pi G  \rho \Delta V,
\end{equation}
where $\Delta V$ denotes the volume of the cell, and $\Delta\bmath{S}$
denotes the vector of the cross-section of the cell
in the direction normal to the cell surface.
We also confirmed that refluxing the gravity is necessary for second-order accuracy by a convergence test (see \S~\ref{sec:fmg_accuracy}).

This refluxing of the gravity is adopted only for the FMG-cycle on the
composite grids shown in Figure~\ref{fmg.eps}{\it a}, in which the
solution converges simultaneously on $\Omega_{\ell}^m$ over the
grid-levels $\ell$.
The refluxing of gravity is not adopted in the V-cycle, since the solution converges sequentially over the grid-levels,

According to the discretization of equation~(\ref{eq:poisson_discrete}), 
the Gauss-Seidel iteration is expressed as,
\begin{equation}
\Phi_{i,j,k}^\mathrm{new} = \Phi_{i,j,k}
-\frac{h^2}{6} R_{i,j,k},
\label{eq:smooth}
\end{equation}
\begin{eqnarray}
R_{i,j,k} &:=&
\partial_x g_{x, i,j,k}
+\partial_y g_{y, i,j,k}
+\partial_z g_{z, i,j,k}
+  4\pi G\rho_{z, i,j,k} \nonumber\\
&=& -{\cal L} \Phi_{i,j,k} +  4\pi G\rho_{z, i,j,k},
\label{eq:residual}
\end{eqnarray}
where $h$ denotes the cell width, $R_{i,j,k}$ denotes the residual,
and ${\cal L}$ denotes the Laplacian operator.
 Equation~(\ref{eq:smooth}) updates 
$\Phi_{i,j,k}$ to $\Phi_{i,j,k}^\mathrm{new}$ at
every iteration. 
Typically only two iterations are required
for the FMG-cycle, and one iteration for the V-cycle.
Hereafter, 
$\Phi_{i,j,k}^\mathrm{new} = {\cal L}^{-1}_\mathrm{GS}\left(
  \Phi_{i,j,k}, \rho_{i,j,k} \right)$ refers to the Gauss-Seidel
iteration given by equations~(\ref{eq:smooth}) and (\ref{eq:residual}).

\subsection{Prolongation and Restriction}

The full-weight prolongation is adopted here, and is given by
\begin{eqnarray}
\Phi^h_{i,j,k} &=& {\cal I}_H^h \Phi^H \nonumber\\
&=& \frac{1}{64}
\Big[
27 \Phi^H_{I,J,K} + \Phi^H_{I\pm1,J\pm1,K\pm1} \nonumber\\
&+& 9 \left(\Phi^H_{I\pm 1,J,K}+\Phi^H_{I,J\pm 1,K}+\Phi^H_{I,J,K\pm 1}\right)\\
&+& 3 \left(\Phi^H_{I,J\pm 1,K\pm 1}+\Phi^H_{I\pm 1,J,K\pm 1}+\Phi^H_{I\pm 1,J\pm 1,K}\right)
\Big], \nonumber
\label{eq:prolongation}
\end{eqnarray}
where $(I,J,K)$ indicates a coarse cell overlapping with a 
fine cell $(i,j,k)$, and the sign of $\pm$ depends on the parity of
 $i$, $j$, and $k$.

The restriction procedure is performed by averaging the 
values of the fine cells $(i,j,k)$ 
which overlap the corresponding coarse cell $(I,J,K)$,
\begin{equation}
\Phi^H_{I,J,K} = {\cal I}_h^H \Phi^h = \frac{1}{8}
\sum_{\hat{i}=i}^{i+1} \sum_{\hat{j}=j}^{j+1} \sum_{\hat{k}=k}^{k+1}
\Phi^h_{\hat{i},\hat{j},\hat{k}}.
\label{eq:restriction}
\end{equation}

The prolongation and restriction introduced above
is used in common with all the
multigrid cycles.

\subsection{FMG-Cycle on AMR hierarchy}

An FMG-cycle based on the standard algorithm of the FMG-cycle for
linear equations (see e.g., \cite{Press91}) is
implemented on the composite grids of the AMR hierarchy,  because the basic operators
are prepared on the composite grids.
The smoothing, prolongation, and restriction procedures are
given by equations~(\ref{eq:smooth}), (\ref{eq:prolongation}), and
(\ref{eq:restriction}), respectively.
In the smoothing procedure, the refluxing of the gravity is performed
according to equation~(\ref{eq:reflux_g}).
In the prolongation, the variables on $\Omega^m_\ell$ 
are transferred onto $\Omega_\ell^{m-1}$. 
Similarly, the restriction procedure transfers variables on
$\Omega^m_\ell$ onto $\Omega^{m+1}_\ell$.
The overlap of coarse grids on fine grids, $\Omega^m_\ell \cap \Omega^m_{\ell+1}$, is therefore not taken into account in the FMG-cycle.  

\subsection{V-Cycle on AMR hierarchy}

In the V-cycle, the multilevel adaptive technique (MLAT) is adopted by
using the full approximation scheme (FAS) (see e.g., \cite{Trottenberg01}).
The solution is iterated towards convergence on $\Omega^m_\ell$, 
where the boundary condition at $\partial \Omega^m_\ell$ is 
obtained from $\Omega^m_{\ell-1}$ according to
equations~(\ref{eq:fmg_boundary1}) and (\ref{eq:fmg_boundary2}).
The prolongation procedure transfers variables from $\Omega_{\ell-1}^m$
to $\Omega_{\ell}^m$,
while the restriction procedure transfers variables 
from $\Omega_\ell^m$ to $\Omega_{\ell-1}^m$.

We now describe the method of the V-cycle in terms of the fine and coarse grids,
$\Omega_{\ell}^m$ and $\Omega_{\ell-1}^m$.
As a pre-smoothing procedure,
the smoothing procedure is applied to 
the initial guess of $\Phi^h$ on $\Omega_{\ell}^m$,
\begin{equation}
\bar{\Phi}^h = {\cal L}^{-1}_\mathrm{GS}\left( \Phi^h, \rho^h \right).
\end{equation}
The density on $\Omega_{\ell-1}^m$ is then obtained as
\begin{equation}
\rho^{H,*} := \left\{
\begin{array}{ll}
{\cal I}^H_h \rho^h + \tau&
\mbox{on } \Omega^m_{\ell-1} \cap \Omega^m_{\ell} \\
\rho^H & \mbox{on the remaining part of } \Omega^m_{\ell-1}
\end{array}
\right. ,
\end{equation}
where
\begin{equation}
\tau  = {\cal L}^H {\cal I}_h^H \bar{\Phi}^h - {\cal I}_h^H {\cal L}^h \bar{\Phi}^h ,
\end{equation}
 is the so-called $\tau$-correction.
The initial guess on $\Omega^m_{\ell-1}$ is given by,
\begin{equation}
\Phi^{H,*} := \left\{
\begin{array}{ll}
{\cal I}^H_h \bar{\Phi}^h &
\mbox{on } \Omega^m_{\ell-1} \cap \Omega^m_{\ell} \\
\Phi^H & \mbox{on the remaining part of } \Omega^m_{\ell-1}
\end{array}
\right. .
\end{equation}
Using $\rho^{H,*}$ and $\Phi^{H,*}$,
the approximate solution $\hat{\Phi}^H$ is obtained from the 
coarser grids.
Then, $\bar{\Phi}^h$ is updated to $\hat{\Phi}^h$ on 
$\Omega^m_\ell$
according to,
\begin{equation}
\hat{\Phi}^h
= \bar{\Phi}^h + {\cal I}^h_H \left( \hat{\Phi}^{H} - {\cal I}^H_h \bar{\Phi}^h \right),
\end{equation}
Finally, post-smoothing is applied to $\hat{\Phi}^h$,
\begin{equation}
\Phi^{h,\mathrm{new}} = {\cal L}^{-1}_\mathrm{GS}\left( \hat{\Phi}^h, \rho^h \right).
\end{equation}
Thus, by this algorithm, the initial guess $\Phi^h$ is updated to $\Phi^{h,\mathrm{new}}$.

\subsection{Utilization of Multigrid Method}

The front-end of the multigrid method described here is the FMG-cycle
on the AMR hierarchy.
Given density $\rho$, the boundary condition for $\Phi$,
and the initial guess $\Phi^0$, 
the FMG-cycle is called iteratively, and
$\Phi^n$ is updated to $\Phi^{n+1}$
by means of the following procedure,
\begin{equation}
R_{i,j,k} = 4 \pi G \rho_{i,j,k} - {\cal L} \Phi^n_{i,j,k} ,
\label{eq:cycle1}
\end{equation}
\begin{equation}
\Phi^{n+1}_{i,j,k} = \Phi^n_{i,j,k} + {\cal L}_\mathrm{FMG}^{-1}
\left(0, R_{i,j,k}\right) ,
\label{eq:cycle2}
\end{equation}
where ${\cal L}_\mathrm{FMG}^{-1}\left(\Phi,\rho\right)$ denotes the
Poisson solver of the FMG-cycle on the AMR hierarchy with initial
guess $\Phi$ and the right-hand side of Poisson's equation $\rho$.
The refluxing of the gravity is implemented in both the ${\cal L}$ and
${\cal L}_\mathrm{FMG}^{-1}$ operators.
The FMG-cycle ${\cal L}_\mathrm{FMG}^{-1}$ is always called for the
fixed boundary condition of zero, because the boundary condition is
transmitted to the residual $R$ according to
equation~(\ref{eq:cycle1}).

The solution converges under the iterative cycle of equations~(\ref{eq:cycle1})
and (\ref{eq:cycle2}), reducing the absolute
value of the residual $|R|$.  In the problem of cloud collapse, the
solution of $\Phi$ in the previous time step is adopted as the initial
guess, and just a single cycle reduces the residual typically by
$\left| R h^2 \right|_\mathrm{max}/\left(4\pi G\rho h^2 \right)_\mathrm{max} \sim 10^{-4}$.

\section{Parallelization and Vectorization}

The code is written entirely in Fortran90, and parallelized by the MPI
library.  The data is partitioned into the computational nodes used for the
parallel calculation in units of the blocks; all cells within a given block are
assigned to the same node.  For a given grid-level, all the blocks
are ordered by means of the Peano-Hilbert space filling curve, and the
blocks are then assigned to the nodes according to this order.
The vectorization in block units is computationally intensive. 
Since a block has $N_x\times N_y\times N_z$ cells, the vector-length
is of the order $N_x\times N_y\times N_z$. \rev{The block sizes,
$N_x$, $N_y$, and $N_z$ are set at the initial configuration, and the
choice of the block size depends on the machine. In the typical case $N_x = N_y = N_z = 8$ is chosen, and the vector-length is therefore
about 512, which is acceptable for the vector processor used
in the calculations. In the next section, convergence tests are performed by
changing the block sizes, $N_x$, $N_y$, and $N_z$.}

\section{Numerical Tests}
\label{sec:test}

\subsection{\rev{Simple Linearized MHD Waves}}

\rev{
Convergence tests for the linearized MHD waves (fast, slow, Alfv\'en,
and entropy waves) are shown.  Following \citet{Crockett05} and
\citet{Gardiner05}, the waves propagation at a slope of 2:1 is performed
on uniform grids and AMR hierarchical grids in the $x-y$ plane.
}

\begin{figure}[t]
\begin{center}
\FigureFile(80m,80mm){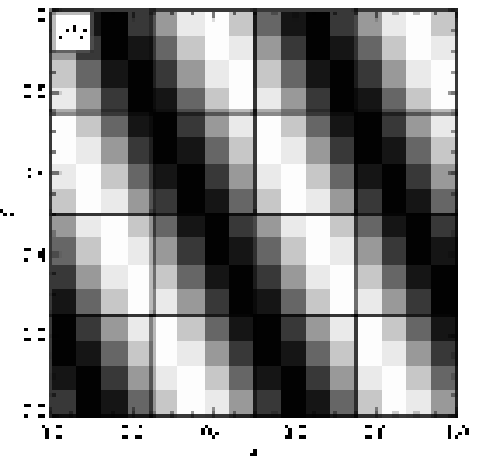}
\FigureFile(80m,80mm){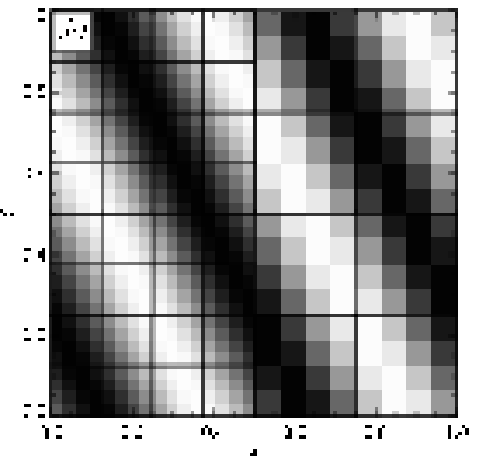}
\end{center}
\caption{ \rev{
Initial conditions for propagation of the simple linearized MHD waves
on ({\it a}) the uniform grid and ({\it b}) the AMR hierarchical grid.
The rectangles denote the block distributions and each block contains $N_x \times N_y = 4^2$ cells.} 
}\label{wave.eps}
\end{figure}

\rev{The unperturbed state is set at, $\rho_0 = 1$, $P_0 = 1$,
$\bmath{v}_0 = 0$, and $\bmath{B}_0 = \sqrt{4\pi}
\left(\frac{1}{\sqrt{2}},\frac{1}{\sqrt{2}},0\right)^T$ in $x,y \in
[0, 1]$.  The specific heat ratio is $\gamma = 5/3$.  All the
linearized MHD waves have wave number of $\bmath{k} = 2\pi (2, 1,
0)^T$.  The wavelength is therefore $\lambda = 1/\sqrt{5}$, and the
computational domain includes two net waves (Fig.~\ref{wave.eps}).
}

\rev{
The initial perturbations of the fast and slow waves are set 
based on the eigenmode, such as,
\begin{eqnarray}
\lefteqn{
\left(
\begin{array}{c}
\delta \rho \\
\delta v_\parallel \\
\delta v_\perp \\
\delta B_\perp \\
\delta P
\end{array}
\right)  } \nonumber\\
&=&
\left(
\begin{array}{c}
\rho_0 \\
c_{F/S} \\
-\frac{B_{\parallel} B_{\perp}}{4 \pi \rho_0}\frac{c_{F/S}}{c_{F/S}^2 - c_a^2} \\
B_{\perp} \frac{c_{F/S}^2}{c_{F/S}^2-c_a^2}\\
(\gamma-1) \left(\rho_0 c_{F/S}^2 - \frac{B_\perp \delta B_\perp}{4 \pi}\right) 
-(\gamma-2)\rho c_s^2
\end{array}
\right)  \nonumber\\
&\times& \delta_\mathrm{pert} \sin (\bmath{k} \cdot \bmath{r}),
\end{eqnarray}
\begin{equation}
c_{F/S}^2 = \frac{1}{2}\left[c_s^2+c_A^2 \pm \sqrt{\left(c_s^2+c_A^2\right)^2-4c_s^2c_a^2 }\right]
\end{equation}
\begin{equation}
c_a^2 = \frac{B_\parallel^2 }{ 4 \pi \rho_0}
\end{equation}
\begin{equation}
c_A^2 = \frac{B_0^2 }{ 4 \pi \rho_0}
\end{equation}
\begin{equation}
c_s^2 = \gamma \frac{P_0}{\rho_0}
\end{equation}
where $v_\parallel$ and $v_\perp$ denote the parallel and
perpendicular components of $\bmath{v}$ with respect to
$\bmath{k}$.  Similarly, $B_\parallel$ and $B_\perp$ denote the
parallel and perpendicular components of $\bmath{B}$ with respect
to $\bmath{k}$.  The subscripts $F$ and $S$ represent the fast and
slow waves.  The amplitude of the perturbation is set at
$\delta_\mathrm{pert} = 10^{-5}$.
For the Alfv\'en wave, the initial perturbation is set as,
\begin{equation}
\left(
\begin{array}{c}
\delta v_z \\
\delta B_z 
\end{array}
\right) 
 = 
\left(
\begin{array}{c}
c_A \\
-B_0 
\end{array}
\right) 
\delta_\mathrm{pert} \sin (\bmath{k} \cdot \bmath{r}).
\end{equation}
The entropy wave is a simple advection wave, and the perturbation is imposed
only on the density, 
\begin{equation}
\delta \rho  = \rho_0
\delta_\mathrm{pert} \sin (\bmath{k} \cdot \bmath{r}),
\end{equation}
and the velocity of the unperturbed state is set to, 
\begin{equation}
\bmath{v}_0 = \frac{\bmath{k}}{k},
\end{equation}
rather than $\bmath{v}_0 = 0$.
}

\begin{figure}[t]
\begin{center}
\FigureFile(80m,80mm){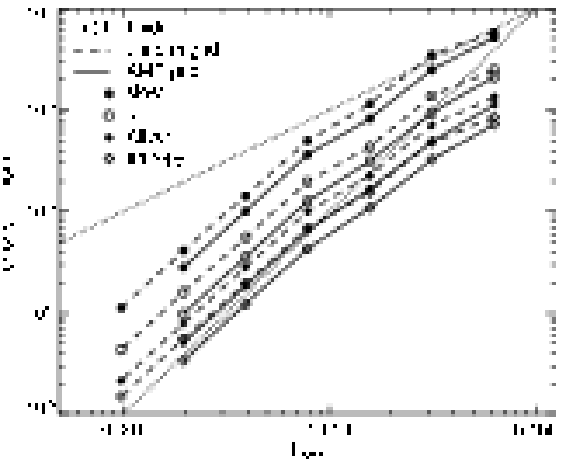}
\FigureFile(80m,80mm){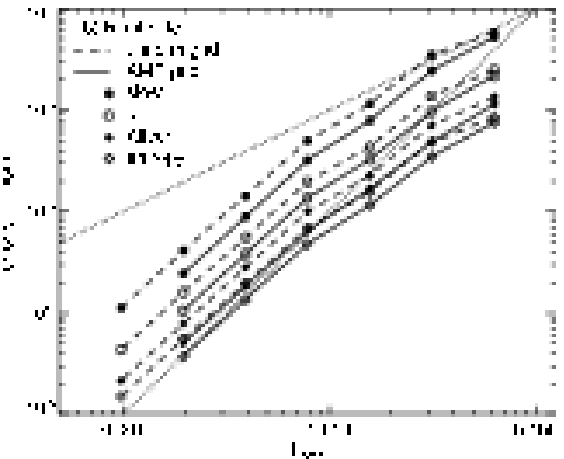}
\end{center}
\caption{ \rev{
$L_1$ norm of error as a function of cell width of the coarsest grid
 $\ell = 0$, $h_\mathrm{base}$, for ({\it a}) the left side ($x \le 0.5$)
and ({\it b}) the right side ($x \ge 0.5$) of the computational domain.
The lines with filled circles, open circles, filled diamonds, and open diamonds
denote errors for slow, fast, Alfv\'en and entropy waves, respectively.
The solid and dashed lines are for the AMR and uniform grids, respectively.
The dotted lines indicate the relationship of errors in proportion to 
$h_\mathrm{base}^2$, and that in proportion to $h_\mathrm{base}$.}
}\label{error.eps}
\end{figure}

\begin{figure}
\begin{center}
\FigureFile(80m,80mm){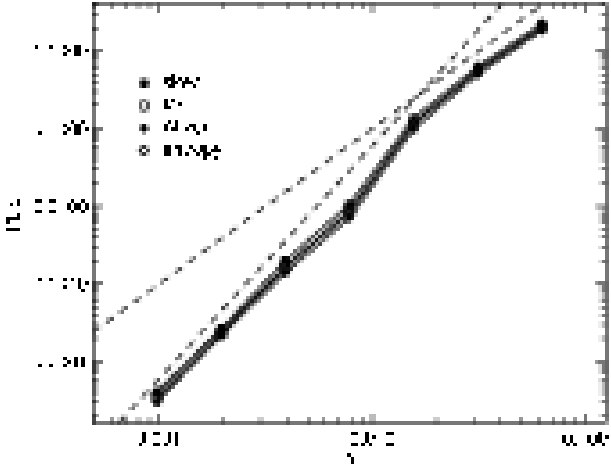}
\end{center}
\caption{ \rev{Decay rate as a function of cell width for slow, fast,
Alfv\'en, and entropy waves in the uniform grid. The lines with
filled circles, open circles, filled diamonds, and open diamonds
denote errors for slow, fast, Alfv\'en and entropy waves, respectively.  The dashed
lines indicate the relationship of errors in proportion to $h^3$, and
that in proportion to $h^2$. } }\label{decayrate.eps}
\end{figure}

\rev{Figure~\ref{wave.eps} shows the wave patterns at the initial
conditions and the block distribution.  The wave propagates in the
upper right direction. The uniform grid consists of $4 \times 4$
blocks of the base grid $\ell = 0$ (Fig.~\ref{wave.eps}{\it a}).  In
the AMR hierarchical grid (Fig.~\ref{wave.eps}{\it b}), the region of
$x \leq 0.5$ is covered by fine grids of $\ell = 1$.  In this
test, the AMR hierarchical grid is static; the block distribution
shown in Figure~\ref{wave.eps} is preserved in the course of the calculations
in order to investigate wave propagation on the coexistence of the
fine and coarse cells.  For both the uniform and AMR grids, the
periodic boundary condition is imposed on the $x = 0, 1$ and $y = 0,
1$.  The three-dimensional code is applied to this problem even though
it has a two-dimensional symmetry, with all the variables being constant in
the $z$-direction.  }

\rev{ We performed the convergence test by changing the cell number
inside a block as $4 \leq N_x , N_y \leq 256$ for the uniform grids,
and $4 \leq N_x , N_y \leq 128$ for the AMR grids.  This corresponds
to a change in the resolution from $h = 1/16$ to 1/1024 for the
uniform grids, from $h = 1/16$ to 1/512 for grid-level $\ell =0$ of
the AMR grids, and from $h = 1/32$ to 1/1024 for grid-level $\ell =1$,
respectively, where $h = \Delta x = \Delta y$.  Each run is halted
when a wave is propagated by a distance of three wavelengths so that
all the waves sweep the entire computational domain.  A $L_1$ norm of
the error is estimated by
\begin{equation}
L_1 = \sum_{i,j,m} \left| U_{i,j}^m - U_{i,j}^{m,0} \right|
\Delta S_{i,j}
\end{equation}
where $\left\{U_{i,j}^m \right\} = \bmath{U}_{i,j}$ denotes the
conservative variables defined by equation~(\ref{eq:Ud}) at the last
stage, $\left\{U_{i,j}^{m,0} \right\} = \bmath{U}_{i,j}^0$ denotes
that of the initial state, and $\Delta S_{i,j} = \Delta x_{i,j} \Delta
y_{i,j}$ denotes the cell surface.  For the case of the AMR grids,
only the cells of $\ell = 1$ are taken into account for the sum in the
region of $x \leq 0.5$ since grid-levels of $\ell = 0$ and 1 
overlap in this region.  }

\rev{ Figure~\ref{error.eps} shows the $L_1$ norm for regions of $x
\leq 0.5 $ and $x \geq 0.5$, separately.  The $L_1$ norms of the
uniform grid (dashed lines) in Figures~\ref{error.eps}{\it a} and
\ref{error.eps}{\it b} coincide with each other because of the
symmetry.  All the waves exhibit second-order accuracy for
$h_\mathrm{base} \lesssim 1/64$, and first-order accuracy for
$h_\mathrm{base} \gtrsim 1/32$, irrespective of the grid types and the
wave modes, where $h_\mathrm{base}$ denotes the cell width of the
coarsest grid (the base grid).  This indicates that the numerical
method basically has second-order accuracy, but the grids of
$h_\mathrm{base} \gtrsim 1/32$ are too coarse to resolve the waves.  }

\rev{ The AMR girds always exhibit smaller $L_1$ norms than the
uniform grids for given a wave mode and $h_\mathrm{base}$.  This is
because the AMR grids resolve the waves using a grid in the
region of $x \leq 0.5$ that is two times finer than the corresponding uniform grids.  This
indicates that the AMR improves the solution of the wave propagation
for all the wave.  However, it should also be noted that the solution
is improved more when the resolution of the uniform grid is increased
by a factor two.  Comparing the left and right sides for the
AMR grids (Figs.~\ref{error.eps}{\it a} and ~\ref{error.eps}{\it
b}), the left region shows slightly smaller errors than the
right region for the fast wave because the left side has finer
resolution.  In contrast, the
left side has slightly larger errors than the right side 
for the slow and entropy waves, in spite of the finer resolution of the left side,
indicating the effects of reflection of the waves at the boundaries
between the fine and coarse grids.  For the Alfv\'en wave, the
significant difference is not observed between the left and right
sides.  }

\rev{ For the AMR girds, a wave propagates from the fine grid into
the coarse grid at $x=0.5$, and it also propagates from the coarse
grid to the fine grid at $x=0$ and 1 via the periodic boundary
condition.  A significant reflection of the waves at $x=0.5$ is
observed in the cases of very coarse grids ($h_\mathrm{base} \gtrsim
1/32$) in the $v_y$ component of the fast wave and the $v_x$ component
of the slow wave.  However, it should be noted that the amplitude of
the components showing reflection is much smaller than that of the
other components, e.g., $v_y$ has an amplitude of only 10~\% of $v_x$
for the fast wave.  }

\rev{ Estimation of decay rate of waves is important for illustrating
properties of a numerical method as indicated by \citet{Crockett05},
and we also estimate the decay rate for the test of the wave
propagation on the uniform grids described above.  The amplitude of
the wave of the $m$-th mode is estimated by
\begin{equation}
w^m = \sum_{i,j} \left| \bmath{l}^{m} (\bmath{U}_{i,j} - \bmath{U}_0) \right|
\end{equation}
where $\bmath{U}_{i,j}$ denotes the conservative variable, 
$\bmath{U}_0$ denotes that in the unperturbed state, and 
$\bmath{l}^{m}$ denotes the left eigenvector of a wave mode
evaluated in the unperturbed state.  The superscript $m$ indicates the
modes of the MHD waves: the fast, slow, Alfv\'en, and entropy waves.
}

\rev{ Figure~\ref{decayrate.eps} shows the decay rates for all the MHD
waves as a function of the cell width.  The decay rates for all the
wave is almost in proportion to $h^3$ for $h \lesssim 1/64$ and $h^2$
for $h \gtrsim 1/64$.  This dependence of the decay rate on the cell
width is consistent with \citet{Crockett05}, indicating that the
region of the second-order accuracy exhibits the third-order accuracy
in the decay rate.  The third-order decay rate of the scheme presented
here is also reported by \citet{Sugimoto04} (see their Fig.~1), in
which the same numerical flux presented here is adopted.  In
contrast, \citet{Ryu95} reported a second-order decay rate by
using a TVD scheme.  The TVD approach is also adopted here but is
based on the scheme of \citet{Fukuda99}, where the MUSCL extrapolation
is applied to the amplitude of the each eigenmode $\bmath{l}^{m}
\bmath{U}$ rather than the primitive variables $\bmath{Q}$.  }

\subsection{\rev{Magnetic Flux Tube}}

\begin{figure*}[t]
\begin{center}
\FigureFile(80m,80mm){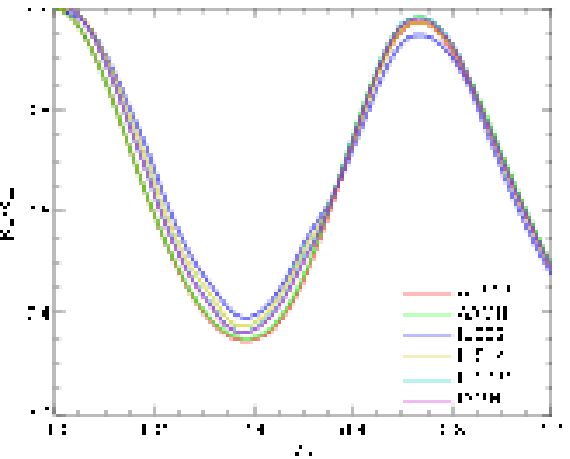}
\FigureFile(80m,80mm){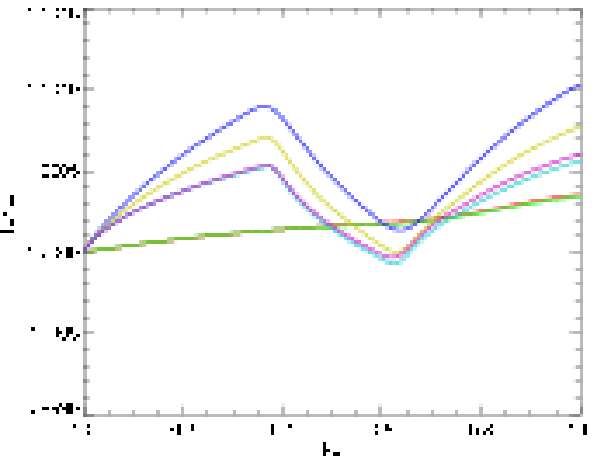}\\
\FigureFile(80m,80mm){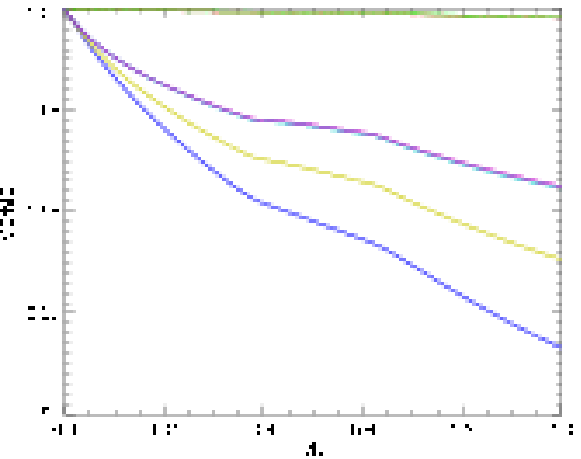}
\FigureFile(80m,80mm){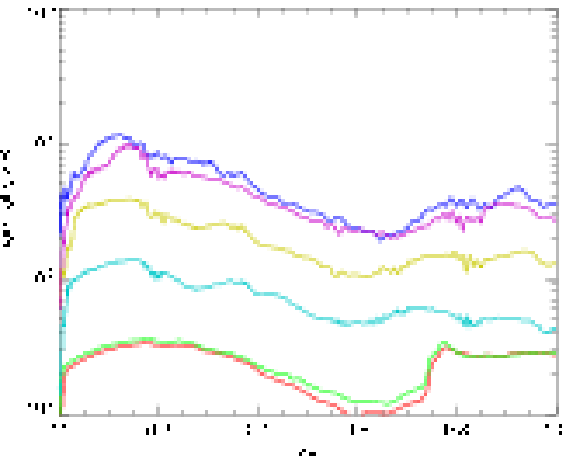}
\end{center}
\caption{ \rev{
Magnetic flux tube problem calculated by various models for 
({\it top left}) kinetic energy, 
({\it top right}) internal thermal energy, 
({\it bottom left}) magnetic energy, 
and ({\it bottom right}) $L_1$ norm of $\nabla \cdot \bmath{B}$.
The abscissa denotes the time normalized by Alfv\'en crossing time
$t_a = 1/c_A = 0.2236$. }
}\label{fluxtube1.eps}
\end{figure*}

\begin{table*}
\caption{Parameters for Magnetic Flux Tube}\label{table:fluxtube}
\begin{center}
\begin{tabular}{llll}
\hline
model & orientation  & cell width ($h$) & grid \\
\hline
\hline
AU512  & aligned  & 1/512              & uniform \\
AAMR   & aligned  & $1/256 - 1/1024$   & AMR ($\ell_\mathrm{max} = 2$) \\
IU256  & inclined & $\sqrt{2}/256$     & uniform \\
IU512  & inclined & $\sqrt{2}/512$     & uniform \\
IU1024 & inclined & $\sqrt{2}/1024$    & uniform \\
IAMR   & inclined & $\sqrt{2}/256 - \sqrt{2}/1024$ & AMR ($\ell_\mathrm{max} = 2$) \\
\hline
\end{tabular}
\end{center}
\end{table*}

\begin{figure*}
\begin{center}
\FigureFile(80m,80mm){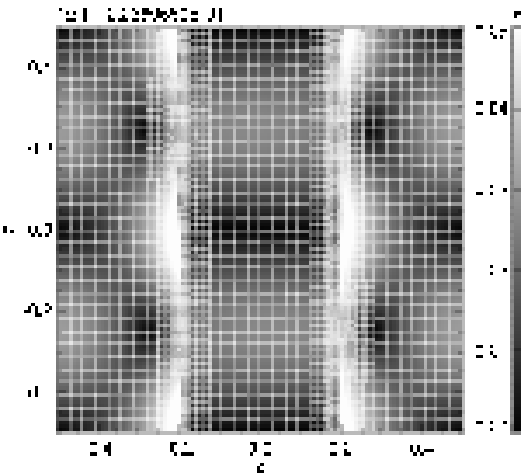}
\FigureFile(80m,80mm){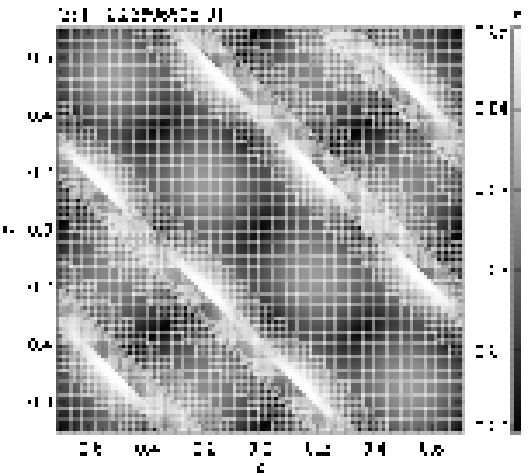}
\end{center}
\caption{ \rev{
Magnetic flux tube for models 
({\it a}) AAMR and ({\it b}) IAMR. 
at the stage of $t/t_a = 1$.
The gray scales denote the velocity $| \bmath{v} |$, and 
the white rectangles denote the block distributions.
({\it a}) The region of $x \in [-0.2, 0.2] $ is magnetized,
and ({\it b}) the diagonal region is non-magnetized. 
The boundaries between the magnetic and non-magnetic regions are
resolved by the fine blocks.
}
}\label{fluxtube.eps}
\end{figure*}

\rev{ The magnetic flux tube problem is proposed by
\citet{Crockett05}, and is a test for stiffness of a MHD scheme.  The
model set-up is the same as that of \citet{Crockett05}.  The
unperturbed state is set as $\rho = 1$, $\bmath{v} = 0$, $B_x =0$, and
$B_z =0 $ in $x,y \in [-0.5, 0.5]$.  The specific heat ratio is
$\gamma = 5/3$.  The computational domain is divided into a magnetized
region of $x \in [-0.2, 0.2]$ and non-magnetized regions.  The
magnetized region and the non-magnetized regions are balanced via the
thermal pressure and magnetic pressure; $B_y =\sqrt{80\pi}$ and $P =
1$ in $x \in [-0.2, 0.2]$, and $B_y =0$ and $P = 11$ in the remaining
regions.  The periodic boundary condition is imposed on the $x = \pm
0.5$ and $y = \pm 0.5$.  On this unperturbed state, the sinusoidal
transverse velocity of $v_x = \delta_\mathrm{pert} c_A \sin(2 \pi y)$
is imposed, and the amplitude of the velocity is set at
$\delta_\mathrm{pert} = 0.01$.  This model is calculated using two types
of grid: a uniform grid of $h = 1/512$ (model AU512), and an AMR grid
(model AAMR), as shown in Table~\ref{table:fluxtube}.  In the AMR
grid, the maximum grid-level is set as $\ell_\mathrm{max} = 2$, and
the following refinement criterion is adopted,
\begin{equation}
\max \left[ {\cal E}(\rho_{i,j,k}), {\cal E}(P_{i,j,k}) \right] \ge 10^{-1},
\label{eq:refinecond1}
\end{equation}
\begin{equation}
{\cal E}(q_{i,j,k}) =  \frac{
\left| \partial^2_x q_{i,j,k} + \partial^2_y q_{i,j,k} \right|
h^2}{q_{i,j,k}} .
\label{eq:refinecond2}
\end{equation}
This criterion captures the curvature of the density and pressure
profiles, and the refinement is attributed mainly to the pressure
jumps in this problem.  Models with uniform grids are
calculated until $t = 6$ $(= 26.83 t_a)$, and the models with the AMR
grid are halted at $t = 0.2236$ $(= t_a)$ because of computational
cost, where $t_a = 1/c_A = 1/\sqrt{20}$ denotes Alfv\'en crossing
time.  }

\begin{figure*}
\begin{center}
\FigureFile(80m,80mm){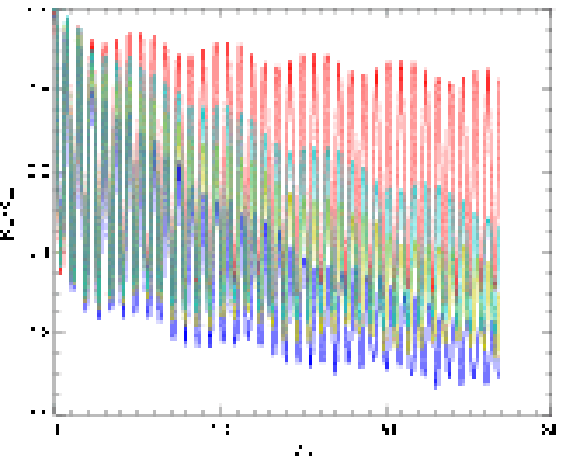}
\FigureFile(80m,80mm){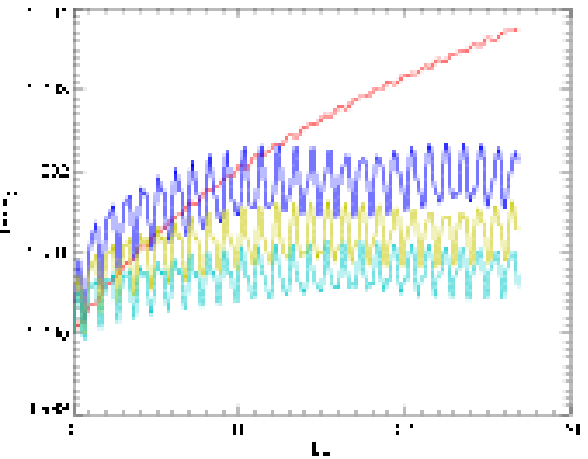}\\
\FigureFile(80m,80mm){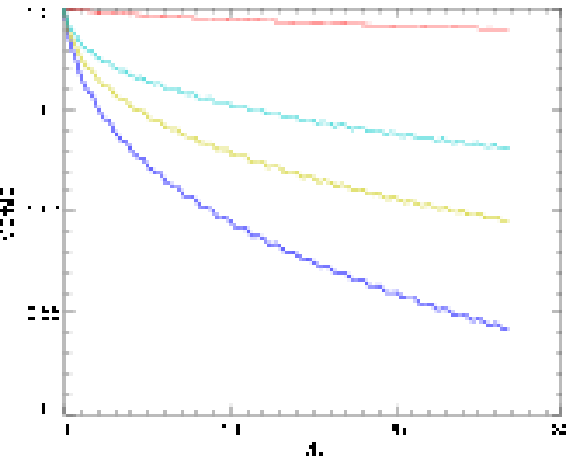}
\FigureFile(80m,80mm){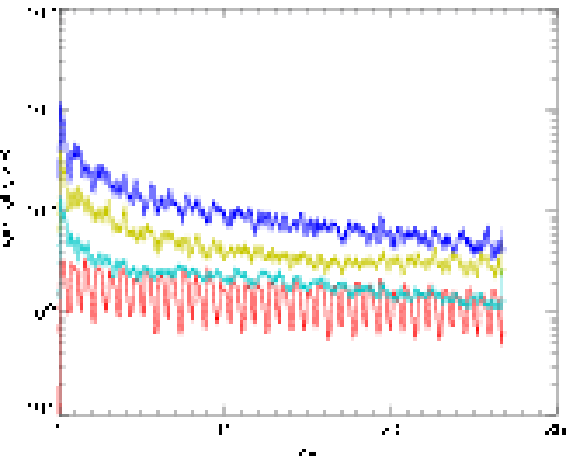}
\end{center}
\caption{ \rev{
Magnetic flux tube problem calculated by the uniform grid models for 
({\it top left}) kinetic energy, 
({\it top right}) internal energy, 
({\it bottom left}) magnetic energy, 
and ({\it bottom right}) $L_1$ norm of $\nabla \cdot \bmath{B}$.
The color legend is same as that of Figure~\ref{fluxtube1.eps}. }
}\label{fluxtube30.eps}
\end{figure*}

\rev{ Grid-inclined flux tube models are also examined.  The model
set-up is the same as that of the grid-aligned model described above, but
the coordinates are rotated at an angle of 45\degree.  Because of this
rotation, the computational box is larger than the grid-aligned model
by a factor $\sqrt{2}$: $x, y \in [1/\sqrt{2},-1/\sqrt{2}]$.  We 
introduce a taper region at the boundaries between the magnetic and
non-magnetic regions, in order to reduce the initial $\nabla \cdot
\bmath{B}$ error.  At the taper region, the pressure and magnetic
field have a $\tanh$ profile, the scale length of which is twice the
cell width.  For the inclined cases, four models are constructed as
shown in Table~\ref{table:fluxtube}; three models are calculated using a
uniform grid by changing the resolution from $h = \sqrt{2}/256$ to
$\sqrt{2}/1024$ (models IU256, IU512, and IU1024), and one using
the AMR grid (model IAMR).} 

\rev{ Figure~\ref{fluxtube1.eps} shows the kinetic energy, internal
thermal energy, magnetic energy, and $L_1$ norm of $\nabla \cdot
\bmath{B}$ error as a function of time for $0 \leq t \leq t_a$. The
three former energies are normalized using the initial values in the
plots.  The divergence error is calculated as follows (see the ninth
component of equations~[\ref{eq:Ud}]--[\ref{eq:Sd}]):
\begin{equation}
\left(\nabla \cdot \bmath{B}\right) _{i,j,k}^{n+1/2}
= \frac{\psi_{i,j,k}^{n} - \psi_{i,j,k}^{n+1} e^{\Delta t c_h^2/c_p^2}}
{c_h^2 \Delta t} .
\label{eq:divB}
\end{equation}
In the period $ 0\leq t \leq t_a$, all the models are calculated
without any indications of the instability.  The two grid-aligned models
(AU512 and AAMR) exhibit good agreement with each other, and $L_1$
norms of the $\nabla \cdot \bmath{B}$ are remain low.  The block
distribution of the AMR models is shown in
Figure~\ref{fluxtube.eps}{\it a}, and the fine blocks cover the
boundaries between the magnetic and non-magnetic regions.  For the
grid-inclined models, the three models (IU256, IU512, and IU1024) are
calculated with uniform grids of different resolutions, and these
models show approximately first-order convergence ($\propto h ^ {1.18}$).
The model with the AMR grid (IAMR) shows an evolution very similar to
the model with the finest uniform grid (IU1024), as shown in
Figrue~\ref{fluxtube1.eps}.  For model IAMR, the boundaries between
the magnetic and non-magnetic regions are resolved by the finest blocks whose
cell width is the same as that of model IU1024
(Fig.~\ref{fluxtube.eps}{\it b}).  This implies that a dominant error
is caused in the boundary between the magnetic and non-magnetic
regions.  However, model IAMR has a $\nabla \cdot \bmath{B}$ error
as large as that of model IU246, which has the same resolution as the coarsest
grid-level of model IAMR.  Examining the velocity
distributions shown in Figure~\ref{fluxtube.eps}, model AAMR exhibits
good agreement with IAMR when it rotates at angle of 45\degree; note
that the diagonal region of Figure~\ref{fluxtube.eps}{\it b}
corresponds to regions of $x\leq -0.2$ and $x \geq 0.2$ in
Figure~\ref{fluxtube.eps}{\it a}.  }

\rev{ Figure~\ref{fluxtube30.eps} is the same as
Figure~\ref{fluxtube1.eps} except for $0 \leq t \leq 6$ and the uniform
grid models.  The long time calculations are performed successfully
without any instabilities for all the uniform gird models.  The
kinetic energy of the grid-aligned model (AU512) decays slightly.  For
the grid-inclined models, the model with higher resolution shows less
decay in the kinetic energy, but the kinetic energies of the finest
grid-inclined model (IU1024) is about half those of the grid-aligned model
(AU512) at the peak of the final osculation.  The internal energies
show different tendencies in the grid-aligned and -inclined models,
but their gradual increases are within the order of 0.1\%.  The $L_1$
norms of the $\nabla \cdot \bmath{B}$ remain low in the grid-aligned
model (model AU512) and the grid-aligned model with the fine grid
(model IU1024).}

\subsection{Double Mach Reflection}
 We consider the double Mach reflection problem as a hydrodynamics test problem.  This test problem was initially proposed by
\citet{Woodward84}, and has since been widely used for testing high resolution
schemes.  A planar shock of Mach 10 travels in a medium of $\rho = 1.4$,
$P=1$, and $\gamma = 1.4$ with incident angle of $60^\circ$ against a rigid wall. The
incident shock interacts with the wall, and a complicated structure
develops featuring a strong and weak reflected shock, contact
discontinuities, and a small jet at the wall (lower boundary; $1/6
\leq x \leq 4$ and $y = 0$).  The computational domain given by $x \in [0,
4]$ and $y \in [0, 1]$ is covered by $32 \times 8$ blocks at $\ell =
0$, and the maximum grid-level is set at $\ell_\mathrm{max} = 4$. 
Each grid has $8^2$ cells ($N_x = N_y = 8$), and so the minimum and
maximum resolutions are $h = 1/64$ and $1/1024$, respectively.
The following refinement criterion is adopted,
\begin{equation}
\max \left[ {\cal E}(\rho_{i,j,k}), {\cal E}(P_{i,j,k}) \right] \ge 10^{-2},
\end{equation}
where ${\cal E}$ is defined by equation~(\ref{eq:refinecond2}).
Three-dimensional code is applied to this problem even
though it has a two-dimensional symmetry, with all the
variables kept constant in the $z$-direction. In order to maintain this
two-dimensional symmetry, the condition $\bmath{F}_z = 0$ is imposed.  The Roe scheme is
modified here based on \citet{Kim03} in order to avoid the carbuncle
instability at shock waves.  
\rev{
The carbuncle instability refers to a protuberant shock profile,
which is an unphysical phenomenon and mashes up shock waves (see, e.g.,
\cite{Perry88}).  
This is thought to be attributed to a low numerical diffusion of the Roe's
scheme, and calculations having high resolutions tend to suffer from this
instability. }

\begin{figure*}
\begin{center}
\FigureFile(80m,80mm){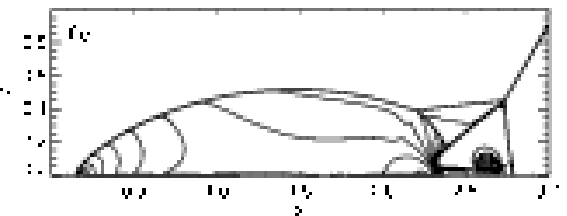}
\FigureFile(80m,80mm){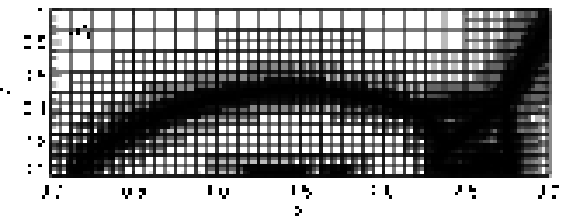}
\FigureFile(80m,80mm){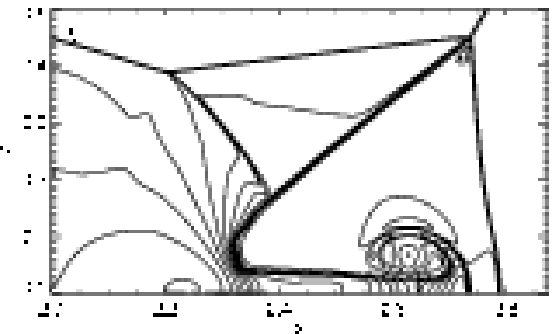}
\FigureFile(80m,80mm){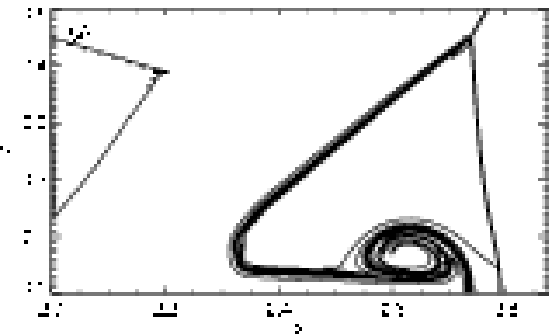}
\end{center}
\caption{ 
Double Mach reflection problem at $t = 0.2$,
({\it a}) density distribution in $[0, 3] \times [0,1]$,
({\it b}) block distribution in $[0, 3] \times [0,1]$,
({\it c}) enlargement of Figure~\ref{MachReflection.eps}{\it a} 
($[2, 2.9] \times [0,0.5]$),
and ({\it d}) distribution of $P/\rho^\gamma$.
In (a) and (c) 30 contour lines are shown in the range 
$1.5 \le \rho \le 22.9705$, while in (d)
 22 contour lines are shown in the range
$1.0 \le P/\rho^\gamma \le 11.5$.
}\label{MachReflection.eps}
\end{figure*}

Figure~\ref{MachReflection.eps}{\it a} shows the density distribution
at $t = 0.2$.  The shock waves are resolved by finest grids as shown in
Figure~\ref{MachReflection.eps}{\it b}.
Figure~\ref{MachReflection.eps}{\it c} shows an enlargement of
Figure~\ref{MachReflection.eps}{\it a} around the double Mach stems.
An eddy is clearly resolved at the end of the jet, and it can be seen even
more clearly in the entropy distribution $P/\rho^\gamma$
(Fig.~\ref{MachReflection.eps}{\it d}).
\rev{
The shock wave is resolved clearly at $x = 2.8$ because  
the carbuncle instability was eliminated.  Without eliminating the carbuncle
instability, a serious instability was observed in this shock wave,
and the eddy was also mashed up.
At $(x, y) = (2.3, 0.1) $, the discontinuity of the sheer begins to
twist, and another eddy may develop as shown by \citet{Shi03}.
}

\rev{
The double Mach reflection problem is also calculated using uniform
grids changing the resolution from $h = 1/64$ to $1/1024$ in order to
confirm the convergence.  This test exhibits an order of convergence of 
1.23, which is slightly higher than the first order, in spite of
implementing the second order accuracy.  This may be because
the slope limiter in the MUSCL extrapolations reduces the accuracy to
the first order in the regions having shock waves and contact
discontinuities. 
}

\subsection{MHD Rotor Problem}

The MHD rotor problem was first proposed by \citet{Balsara99}, and it tests
the propagation of non-linear Alfv\'en waves.  The model set-up is in the 
same manner as that of \citet{Toth00} and \citet{Crockett05}.  The
computational domain is a unit square of $x,y \in [0, 1]$.  In the
initial stage, a uniform cylinder with $\rho = 10$, $P=1$, and
radius 0.1 rotates at an angular velocity of $20$.  The ambient
medium is at rest with $\rho = 1$, $P=1$, and $\bmath{v}=0$.  At the
boundary between the cylinder and ambient medium, a taper region is
used in order to reduce the initial transition (see \cite{Toth00}).
The computational domain is subject to a uniform magnetic field, 
$(B_x, B_y, B_z) = (5, 0, 0)$. The adiabatic index of the gas is
$\gamma = 1.4$.
The computational domain is covered by $16\times16$ blocks at 
$\ell = 0$, and each block consists of $8\times8$ cells ($N_x = N_y = 8$).
The maximum grid-level is set at $\ell_\mathrm{max}=2$.
The coarsest and finest grids therefore exhibit an effective resolution of
$h=1/128$ and $1/512$, respectively. 
The following refinement criterion is adopted,  
\begin{equation}
\mathrm{max}\left[ {\cal E}(\rho_{i,j,k}), {\cal E}(P_{i,j,k})
\right] \ge 1,
\end{equation}
\begin{equation}
{\cal E}(q_{i,j,k}) = \left[ {\cal E}_x^2(q_{i,j,k})
+{\cal E}_y^2(q_{i,j,k}) +  {\cal E}_z^2(q_{i,j,k})\right]^{1/2} ,
\end{equation}
\begin{eqnarray}
\lefteqn{
{\cal E}_x(q_{i,j,k})
} \nonumber \\
&=&  \frac{ h^2 \partial_{x}^2 q_{i,j,k} }
{ h \left| \partial_x q_{i+1/2,j,k} \right| 
+ h \left| \partial_x q_{i-1/2,j,k} \right| 
   + \epsilon  h^2 \partial_{x}^2 q_{i,j,k}},
\end{eqnarray}
and ${\cal E}_y$ and ${\cal E}_z$ are defined in a similar manner to
${\cal E}_x$, where
$\epsilon = 10^{-2}$ (e.g., \cite{Fryxell00}).

\begin{figure*}
\begin{center}
\FigureFile(160m,160mm){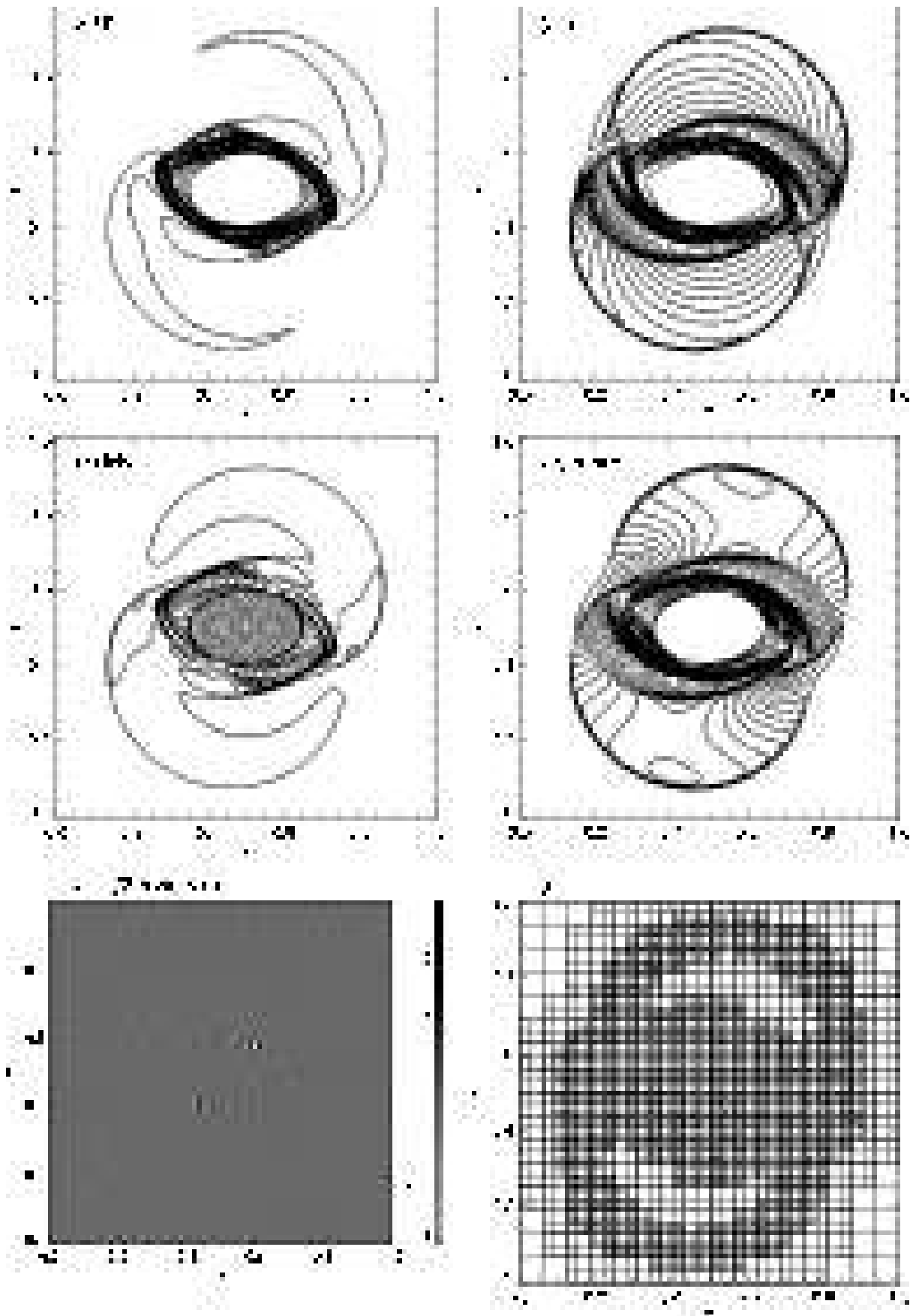}
\end{center}
\caption{ 
MHD rotor problem solved by the AMR code.
({\it a}) Density, ({\it b}) thermal pressure, ({\it c}) Mach number,
({\it d}) magnetic pressure, 
({\it e}) magnetic monopole, and
({\it f}) grid distribution are shown at $t=0.15$.
In (a)-(d) 30 contour lines are shown in the ranges of 
$0.483 \le \rho \le 12.95$,
$0.0202 \le P \le 2.008$,
$0 \le |\bmath{v}|/c_s \le 8.18$,
and 
$0.0177 \le \bmath{B}^2/8\pi \le 2.642$, respectively.
}\label{MHD_rotor.eps}
\end{figure*}

Figures~\ref{MHD_rotor.eps}{\it a}-\ref{MHD_rotor.eps}{\it d} show the density, thermal
pressure, Mach number, magnetic pressure at $t=0.15$.  The finest grids
capture the outer shock
fronts and inner complex structures, as shown in Figure~\ref{MHD_rotor.eps}{\it f}.  
The distribution of the physical
variables plotted here exhibit excellent agreement with Figure~18 of
\citet{Toth00} and Figure~12 of \citet{Crockett05}.
Figure~\ref{MHD_rotor.eps}{\it e} shows the amplitude of the magnetic
divergence error normalized by the cell width and the local
magnetic field strength, $h (\nabla \cdot \bmath{B}) / |\bmath{B}|$,
where $\nabla \cdot \bmath{B}$ is estimated using equation~(\ref{eq:divB}).
The divergence error reaches a maximum value of $1.5\times10^{-2}$ at 
the inner discontinuity, and is $2\times10^{-3}$ at the outer shock fronts.
The divergence cleaning of \citet{Dedner02} keeps the divergence error small. 

\subsection{Accuracy of Multigrid Method}
\label{sec:fmg_accuracy}

The accuracy of the multigrid method is examined using the approach of \citet{Matsumoto03}.
Two uniform spheres of masses 1 and 2, both having a radius of 6/1024,
 are located 
  at $(x,y,z)=(12/1024, 0, 0)$ and
$(-12/1024, 0, 0)$
in 
the computational domain $x, y, z \in [-0.5, 0.5]$.
 A Dirichlet physical boundary condition is imposed
on $\Phi$, and it is evaluated by 
the multi-pole expansion of the density distribution at $\ell=0$, where
the monopole, dipole, and quadruple moments are taken into account.

The computation was started from the initial guess $\Phi = 0$.  The
residual decreased by more than a factor of several hundred with each
FMG iteration given by equations~(\ref{eq:cycle1}) and
(\ref{eq:cycle2}).  After 10 iterations, the residual was of the order
of the round-off error. The numerical solution for the gravity obtained by this computation is compared to the exact analytic solution.

\begin{figure}
\begin{center}
\FigureFile(80m,80mm){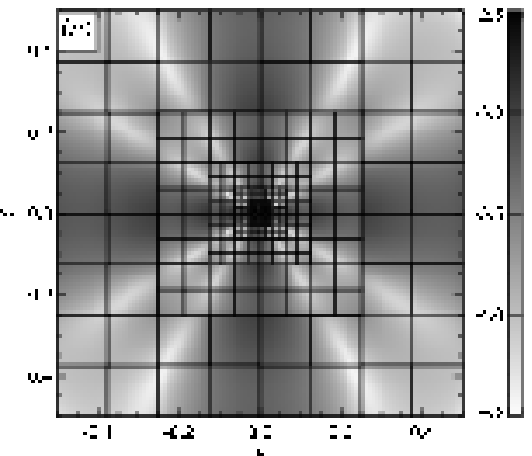}
\FigureFile(80m,80mm){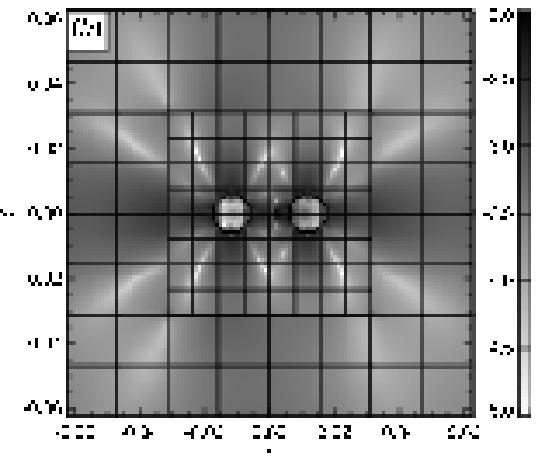}
\end{center}
\caption{ 
Distribution of the relative error of the numerically calculated gravity in the $z=0$ plane with $N_x = N_y =
N_z =16$ after 10 iterations of the FMG-cycles.
The gray scale denotes $\log \left( \left| \bmath{g} - \bmath{g}_\mathrm{ex}\right|/ 
\left| \bmath{g}_\mathrm{ex} \right| \right)$, and
the rectangles indicate the block distribution.  Panel {\it b} is an enlargement
of panel {\it a}.  
}\label{fmg_error_dist.eps}
\end{figure}

Figure~\ref{fmg_error_dist.eps} shows the relative error of
the numerically calculated gravity,
$\left| \bmath{g} - \bmath{g}_\mathrm{ex}\right|/ 
\left| \bmath{g}_\mathrm{ex} \right|$,
on a logarithmic scale, where $\bmath{g}_\mathrm{ex}$ denotes the exact
gravity obtained analytically. 
In this calculation, $N_x=N_y=N_z =16$ and
$\ell_\mathrm{max} = 4$ are adopted. The block distribution is 
also shown in Figure~\ref{fmg_error_dist.eps}.
The maximum error occurs at the edges of the spheres due to
the discretization error that arises from the sharp density contrast of the spheres.
In the other domains, the error is less than $\sim 10^{-3}$.
It is noteworthy that no significant error appears at the
interfaces between the coarse and fine grids.  This is attributed to
the refluxing of the gravity described in \S~\ref{sec:smoothing}.

\begin{figure}
\begin{center}
\FigureFile(80m,80mm){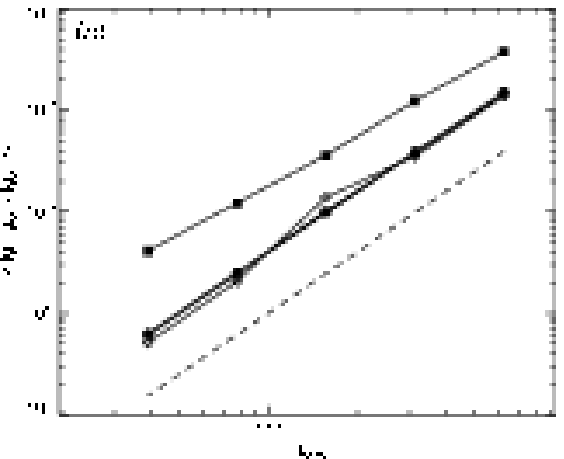}
\FigureFile(80m,80mm){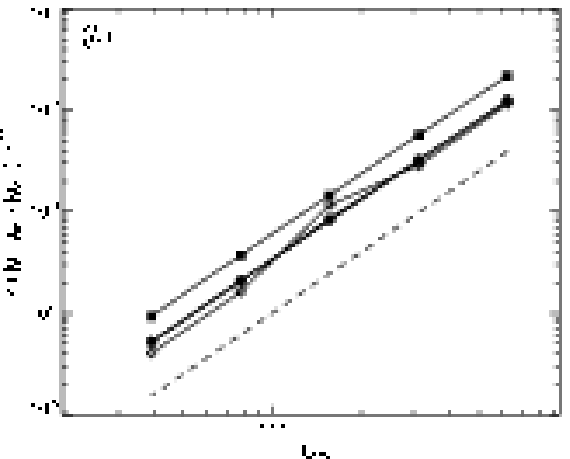}
\FigureFile(80m,80mm){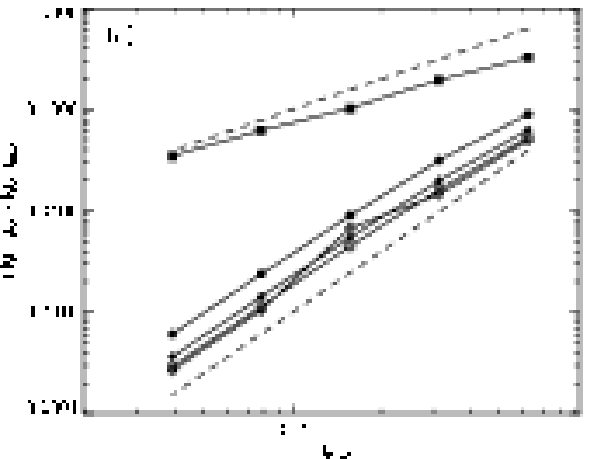}
\end{center}
\caption{ 
The relative error of the numerically computed gravity $ \left| \bmath{g} - \bmath{g}_\mathrm{ex}
\right| / \left| \bmath{g}_\mathrm{ex} \right|$ as a function of
$h_\mathrm{base}$ (cell width of the coarsest grids). 
The error is measured by ({\it a}) average, ({\it b}) root mean
square, and ({\it c}) maximum on each grid-level separately.
The open diamonds, open circles, filled diamonds, filled circles, and
filled square denote the errors on grids of 
$\ell = 0, 1, 2, 3$, and 4, respectively.
 The dashed lines indicate the relationship of 
({\it a}-{\it c}) errors in proportion to $h_\mathrm{base}^2$, 
and ({\it c}) those in proportion to $h_\mathrm{base}$
}\label{fmg_error.eps}
\end{figure}

Figure~\ref{fmg_error.eps} shows the dependence of the error on resolution.
The error of the gravity is measured by changing the 
number of cells inside a block, $N_x$, $N_y$, and
$N_z$, but fixing the block distribution.
The ordinates of Figures~\ref{fmg_error.eps}{\it a},
\ref{fmg_error.eps}{\it b}, and \ref{fmg_error.eps}{\it c} denote
the errors measured by the average, root mean square, and maximum
values, corresponding to the $L_1$, $L_2$, and $L_\infty$ norms, respectively.
The abscissa denotes the cell width of the coarsest grid $\ell = 0$.
The five lines show the errors at the grid-levels of $\ell = 0, \cdots, 4$, while 
the five points in the lines denote the errors for $N_x = N_y = N_z = 2$,
4, 8, 16, and 32.
All the lines, except for $\ell = 4$ in Figure~\ref{fmg_error.eps}{\it c},
exhibit second-order accuracy for the multigrid method presented here. 
By contrast, the line of $\ell = 4$ in Figure~\ref{fmg_error.eps}{\it c} shows a first-order accuracy. This is attributed to the 
discretization error of the density imposed on this grid-level.

\subsection{Convergence of Multigrid Method}
\label{sec:fmg_conv}

\begin{figure}
\begin{center}
\FigureFile(80m,80mm){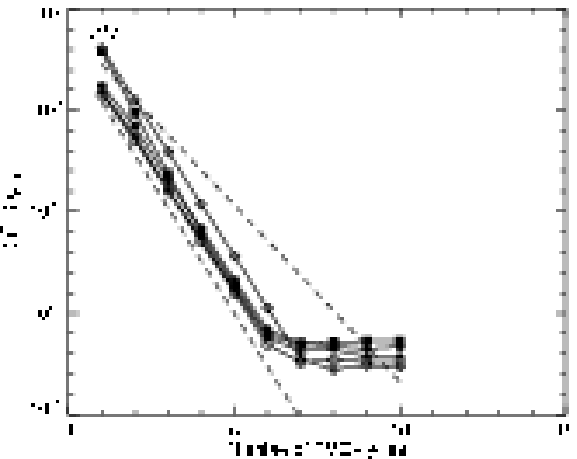}
\FigureFile(80m,80mm){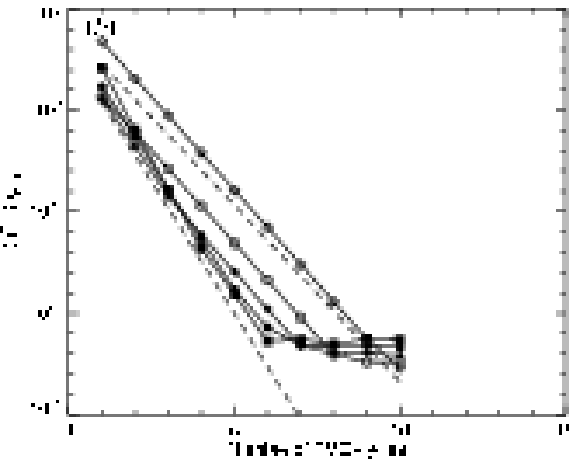}
\FigureFile(80m,80mm){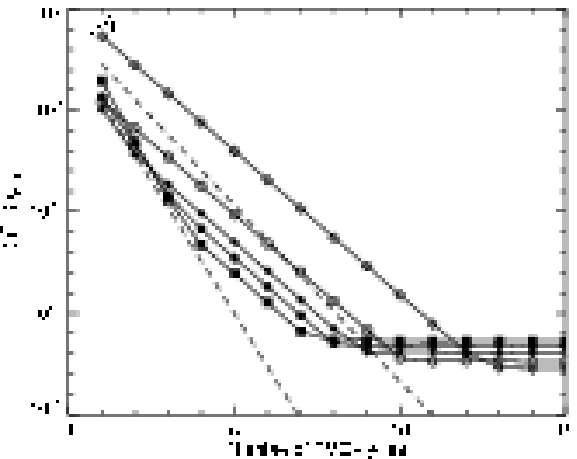}
\end{center}
\caption{ 
The maximum residual ($|h^2 R|_\mathrm{max}$) as a function of the
number of the FMG-cycles for the cases
({\it a}) $N_x = N_y = N_z = 8$,
({\it b}) $N_x = N_y = N_z = 16$,
and ({\it c}) $N_x = N_y = N_z = 32$.
The open diamonds, open circles, filled diamonds, filled circles, and
filled square denote the residuals measured on grids of 
$\ell = 0, 1, 2, 3$, and 4, respectively.
The dashed line displays the relationships 
$\left|h^2 R^n\right|_\mathrm{max} \propto \exp\left( -4 n \right)$ and $\exp \left(-6 n \right)$,
where $n$ denotes the number of the FMG-cycles.
}\label{fmg_converge_L.eps}
\end{figure}

The reduction of the residual defined by equation~(\ref{eq:residual}),
is measured to evaluate the efficiency of the iteration for the same
problem as described in \S~\ref{sec:fmg_accuracy}.
Figure~\ref{fmg_converge_L.eps} shows the maximum residual in each
grid-level as a function of the number of the FMG-cycles defined by
equations~(\ref{eq:cycle1}) and (\ref{eq:cycle2}) for the cases
$N_x = N_y = N_z =$ 8, 16, and 32.  The residuals plotted here are
multiplied by $h^2$ so that they have a dimension of $\Psi$.  For the
case $N_x = N_y = N_z = 8$ (Fig.~\ref{fmg_converge_L.eps}{\it a}),
the residuals in all the grid-levels decrease in proportion to
$\exp(-6n)$.  After 7 iterations of the FMG-cycles, the residuals at all
grid-levels reach the round-off error.  On the other hand, for the
cases $N_x = N_y = N_z = 16$ and 32
(Figs.~\ref{fmg_converge_L.eps}{\it b} and \ref{fmg_converge_L.eps}{\it
  c}), the rate at which the residuals decrease on the coarse grid-levels are
slower than those of the fine grid-levels. 
The slower convergence is due to the boundary condition imposed on the
coarsest grid.

\subsection{Collapse and Fragmentation of an Isothermal Cloud}

  The present numerical technique is applied to the problem of the collapse
and fragmentation of an isothermal cloud as a gravitational hydrodynamics test problem. 
While isothermal collapse has been calculated by many authors,
the particular model given in \citet{Bate97} and \citet{Truelove98} is followed here.
The initial cloud has a uniform, spherical density distribution, and
 rotates rigidly around the $z$-axis.
The mass of the cloud is $1 M_\odot$, and the radius is $R_c = 5\times
10^{16} \,\mathrm{cm}$, and so  the density of the cloud is therefore
$\rho_0 = 3.79\times10^{-18}\,\mathrm{g}\,\mathrm{cm}^{-3}$. 
Conventionally, a cloud is
characterized by two global quantities, 
$\alpha = E_\mathrm{th}/ |E_\mathrm{grav}|$
and $\beta = E_\mathrm{rot}/ |E_\mathrm{grav}|$, where
$E_\mathrm{th}$, $E_\mathrm{rot}$, and $E_\mathrm{grav}$ denote the
thermal, rotation, and gravitational energies.
The cloud here has $\alpha = 0.26$ and $\beta = 0.16$.
The sound speed and the angular velocity are obtained as
$c_s = 0.166\,\mathrm{km}\,\mathrm{s}^{-1}$ and
$\Omega = 7.14\times 10^{-13}\,\mathrm{s}^{-1}$, respectively.
The cloud is perturbed by a bar perturbation with an amplitude of 10\%;
$\rho = \rho_0 (1+0.1 \cos 2 \phi)$.
The cloud is embedded in an ambient gas, whose density is $0.01\rho_0$.

The computational domain is
$x, y, z \in [-2R_c, 2R_c,]\times [-2R_c, 2R_c,]\times [0, 2R_c,]$,
and a mirror boundary condition is imposed on the $z = 0$ plane.
Blocks of $N_x = N_y = N_z = 8$ are adopted.
At the initial stage, $16\times16\times8$ blocks are distributed 
at the grid-level of $\ell =0$. 
The cloud radius $R_c$ is therefore resolved by 32 cells.
This initial resolution is same as that of \citet{Truelove98}.
The Jeans condition is employed as a refinement criterion;
the blocks are refined when the Jeans length is shorter than 8 times of cell width;
$(\pi c_s^2/G\rho)^{1/2} < 8 h$.
This refinement criterion is twice as severe as that of \citet{Truelove97}.

\begin{figure}
\begin{center}
\FigureFile(80m,80mm){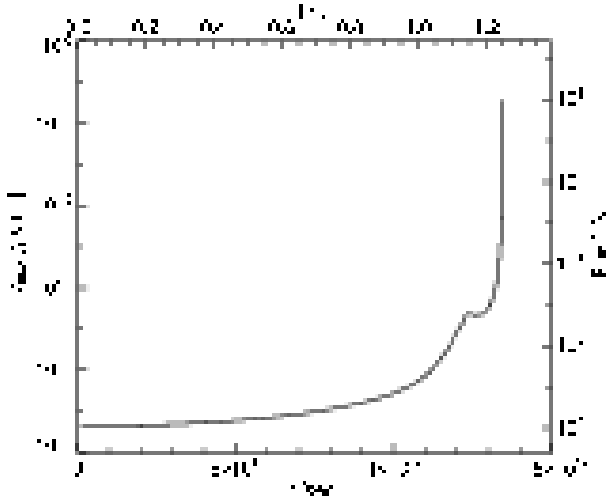}
\end{center}
\caption{ 
Maximum density as a function of time for the problem of the
collapse and fragmentation of an isothermal cloud.
}\label{rhomax.eps}
\end{figure}

Figure~\ref{rhomax.eps} shows the maximum density as a function of time. The central density initially increases gradually as the cloud collapses.  
A thin disk forms due to the fast rotation of the cloud when the central density reaches a value of around
$\rho_\mathrm{max} = 2\times10^{-15}\,\mathrm{g}\,\mathrm{cm}^{-3}$ at
$t = 1.25\times 10^{12}\,\mathrm{s}$ ($=1.16\,t_\mathrm{ff}$),
where $t_\mathrm{ff} = (3 \pi / 32 G \rho_0)^{1/2} = 5.60\times
10^{12} \,\mathrm{s}$, and this disk bounces in the $z$-direction. After the bounce, 
the density increases more rapidly, and approaches a singularity at
$t=1.35\times10^{12}\,\mathrm{s}$ ($=1.25\,t_\mathrm{ff}$).

\begin{figure}
\begin{center}
\FigureFile(80m,80mm){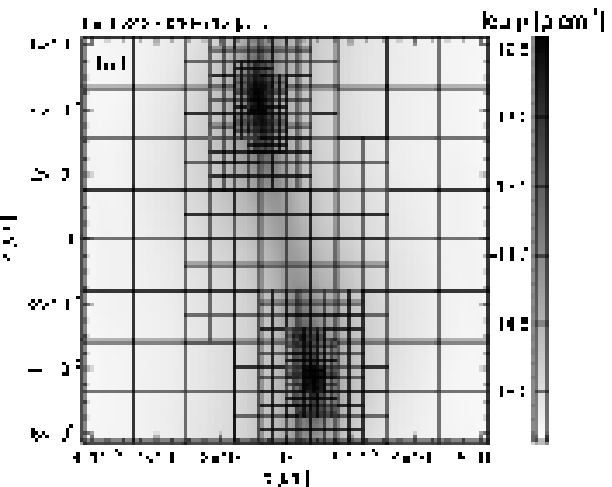}
\FigureFile(80m,80mm){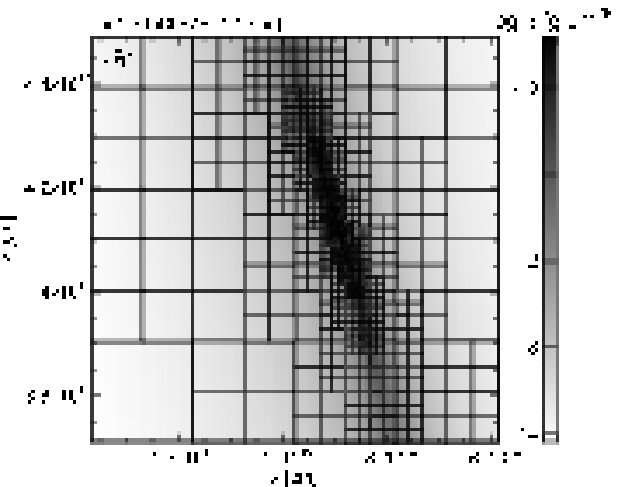}
\end{center}
\caption{ 
Collapse and fragmentation of a rotating isothermal cloud.
The gray scale denotes logarithmic density in the $z=0$ plane, and
the rectangles indicate the block distribution. 
({\it a}) The central region of the cloud  when
$\rho_\mathrm{max} = 1.02 \times 10^5 \rho_0$ on the grid-levels $\ell = 3-7$.  
({\it b}) 
Enlargement around the upper left fragment when
$\rho_\mathrm{max} = 1.01 \times 10^8 \rho_0$ on   grid-levels $\ell = 6-12$.  
}\label{fragment.eps}
\end{figure}

Figure~\ref{fragment.eps}{\it a} shows the density distribution in the
$z=0$ plane when $\rho_\mathrm{max} = 1.02 \times 10^5 \rho_0$.
The disk has two density peaks, which are resolved by the fine blocks.
Each density peak has a filamentary structure.
This density structure was also found by \citet{Truelove98} (see
their Fig.~12).  

Figure~\ref{fragment.eps}{\it b} shows the upper left bar when$\rho_\mathrm{max} = 1.01 \times 10^8 \rho_0$.  The bar collapses
to form a very narrow filament.  This is also consistent with
\citet{Truelove98} (see their Fig.~13).

\subsection{Collapse and Outflow Formation of a Cloud Core with Slow Rotation and Oblique
  Magnetic Field}

The collapse of a cloud with magnetic field parallel to the rotation axis
has been simulated by
\citet{Machida04,Machida05a,Machida05b,Zieglar05,Banerjee06,Fromang06}, while the
collapse of an oblique magnetic field has been simulated by
\citet{Matsumoto04a,Machida06}.  In the case of an oblique magnetic
field, the rotation axis changes its direction due to the anisotropic
magnetic braking during the collapse, and an outflow is ejected
after an adiabatic core formation. 

In this paper, the model MF45 of \citet{Matsumoto04a} was calculated using
the AMR presented here.  The initial cloud has the density
profile of a critical Bonnor-Ebert sphere
\citep{Ebert1955,Bonnor1956}, but the density is increased by a factor of 1.68.
The central density is $\rho_0 = 1 \times
10^{-19}\,\mathrm{g}\,\mathrm{cm}^{-3}$, and the radius of the cloud
is $R_c = 5.49\times 10^{17}\,\mathrm{cm}$.
The cloud rotates slowly at an angular velocity of 
$\Omega_0 =7.11\times10^{-7}\,{\rm yr}^{-1} (= 0.15 t_{\rm ff}^{-1})$,
where the freefall time is $t_{\rm ff} = 2.10\times10^5\,{\rm yr} $.
The initial magnetic field is inclined at an angle of $45^\circ$
with respective to the $z$-axis (rotation axis), and has a strength 18.6~$\mu$G.
The cloud has parameters given by
$\alpha=0.5$, $\beta = 0.02$, and $E_\mathrm{mag}/|E_\mathrm{grav}| =
0.721$, where $E_\mathrm{mag}$ denotes the total magnetic energy
inside the cloud.

In order to mimic the formation of the first stellar core, the equation of
state is changed according to the density: an isothermal gas of
$T=10$~K ($c_s = 0.19\,\mathrm{km}\,\mathrm{s}^{-1}$) is assumed when
density is less than $\rho_\mathrm{cr} = 2\times
10^{-13}\,\mathrm{g}\,\mathrm{cm}^{-3}$, while a polytrope gas of
$\gamma = 5/3$ is assumed when density is higher than $\rho_\mathrm{cr}$.

\begin{figure}[t]
\begin{center}
\FigureFile(80m,80mm){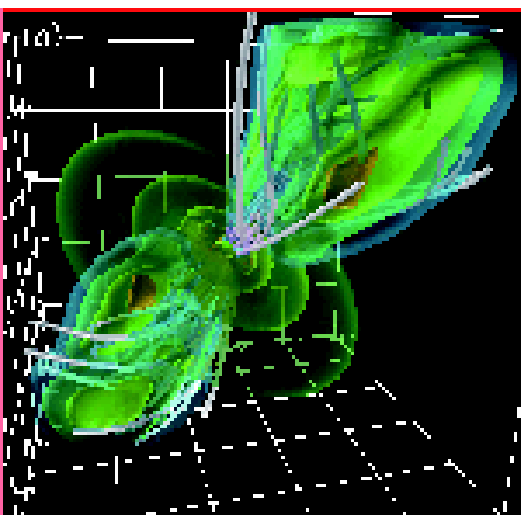}
\FigureFile(80m,80mm){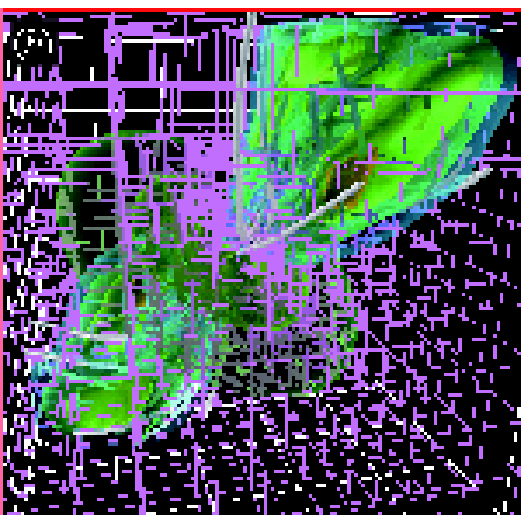}
\end{center}
\caption{ 
({\it a}) Three-dimensional view of protostellar collapse and outflow
  formation at $t = 4.81 \times 10^5$~yr.
The 6 disk-shaped isosurfaces are shown for 
$4.7 \leq \log \rho/\rho_0 \leq 7.0$, and 
the 4 bi-polar isosurfaces are shown for 
$2.0 \leq v_r/c_s \leq 9.0$.
The tubes indicate the magnetic field lines.
The coordinates are shown in the units of $c_s/(4\pi G \rho_0)^{1/2}$.
\rev{({\it b}) Same as panel {\it a} but the block distribution is overplayed.
The grid-levels is ranged from 11 to 13.}
}\label{outflow.eps}
\end{figure}

Figure~\ref{outflow.eps}{\it a} shows the central region of (287~AU)$^3$
at $t = 4.81 \times 10^5$~yr.
As calculated by \citet{Matsumoto04a}, the outflow is ejected from the
region near the first core.  The outflow speed reaches $10 c_s$, which
is also consistent with a previous calculation. 

Figure~\ref{outflow.eps}{\it b} shows the block distribution, where
 grid-levels of $\ell = 11 - 13$ are shown.  Each grid-level has
$8^3$ blocks.  This grid distribution reproduces effectively 
the same cell distribution as a nested grid including $64^3$ cells
in each grid-level.

\section{Summary}
\label{sec:summary}

SFUMATO, a self-gravitational MHD code applying the AMR technique is presented. 
The grid is configured in a block structure.  

The MHD scheme is implemented so that it is of second-order spatial
 accuracy by means of the TVD approach.  The upwind numerical
flux is obtained by the linearized Riemann solver.  The scheme is
fully cell-centered, and the divergence error of the magnetic fields is
cleaned.
\rev{The convergence tests of the linearized MHD waves exhibit a
 second-order accuracy, and the decay rates shows a third-order
 accuracy.  Stiffness of the scheme is confirmed by the 
MHD flux tube problems.}

The self-gravity is solved by a multigrid method composed of:
(1) a FMG-cycle on the AMR hierarchical grids, (2) a V-cycle on these
grids, and (3) a FMG-cycle on the base grid.  The FMG-cycle on the AMR
hierarchical grids enables a scalable dependence on the number of cells.
The multigrid method ensures that the solution converges rapidly; the residual is reduced
by a factor of $10^{-3}-10^{-2}$ every iteration.  The multigrid method
exhibits a second-order spatial accuracy.  Moreover, no spurious
features appear at the interfaces between fine and coarse grid-levels,
due to the flux conservation at the interface.

The MHD scheme and multigrid method are combined so as to have
 second-order temporal accuracy.  The time-marching has two
modes: synchronous and adaptive time-step modes. The former mode is
adopted for problems including self-gravity, while the latter mode is used
for all other problems.

The AMR code was tested by considering test problems given by double Mach
reflection, the MHD rotor, fragmentation of an isothermal cloud, and
outflow formation in a collapsing cloud.  The present results are in good agreement with those of previous studies conducted by the present author as well as other authors.

\bigskip

The author would like to thank K. Tomisaka and T. Hanawa for valuable discussions.
The author thanks the anonymous referee for his/her valuable comments,
which have been very helpful in improving the paper. 
Numerical computations were carried out on VPP5000 at the
Center for Computational Astrophysics, CfCA, of the National
Astronomical Observatory of Japan, 
and SX6 at Japan Aerospace Exploration Agency, JAXA.
This research was supported in part 
by the Hosei Society of Humanity and Environment, 
and by Grants-in-Aid 
for Young Scientists (B) 16740115, 18740113,
for Scientific Research (C) 17540212,
and for Scientific Research (B) 17340059
awarded by the Ministry of Education, Culture, Sports, Science and Technology, Japan.

\end{document}